\documentclass[11pt]{article}

\newcommand{\blind}{0}

\usepackage{amsmath}
\usepackage{graphicx}
\usepackage{enumerate}
\usepackage{natbib} 
\usepackage{url} 

\graphicspath{{figures/}}
\usepackage{amsfonts, amsthm, latexsym, amssymb}
\usepackage{dsfont}
\usepackage{array}
\usepackage{float}
\usepackage{multirow}
\usepackage{xcolor}
\usepackage{lineno}
\usepackage[ruled,vlined]{algorithm2e}
\newcommand{\mb}{\mathbf}
\newcommand{\bs}{\boldsymbol}
\newcommand{\mc}{\mathcal}

\begin{document}

\def\spacingset#1{\renewcommand{\baselinestretch}%
{#1}\small\normalsize} \spacingset{1}

\if0\blind
{
\title{\bf The temporal overfitting problem with applications in wind power curve modeling}
\author{Abhinav Prakash, Rui Tuo, and Yu Ding \\
	Department of Industrial and Systems Engineering\\
	Texas A\&M University}
\date{}
\maketitle
} \fi

\if1\blind
{			
\title{\bf The temporal overfitting problem with applications in wind power curve modeling}
\author{}
\date{}	
\maketitle
\medskip
} \fi

\begin{abstract}
This paper is concerned with a nonparametric regression problem in which the input variables and the errors are autocorrelated in time. The motivation for the research stems from modeling wind power curves. Using existing model selection methods, like cross validation, results in model overfitting in presence of temporal autocorrelation. This phenomenon is referred to as temporal overfitting, which causes loss of performance while predicting responses for a time domain different from the training time domain.  We propose a Gaussian process (GP)-based method to tackle the temporal overfitting problem. Our model is partitioned into two parts---a time-invariant component and a time-varying component, each of which is modeled through a GP.  We modify the inference method to a thinning-based strategy, an idea borrowed from Markov chain Monte Carlo sampling, to overcome temporal overfitting and estimate the  time-invariant component. We extensively compare our proposed method with both existing power curve models and available ideas for handling temporal overfitting on real wind turbine datasets. Our approach yields significant improvement when predicting response for a time period different from the training time period. Supplementary material and computer code for this article is available online.
\end{abstract}

\noindent%

\spacingset{1.5} 

\section{Introduction}\label{Section:Intro}

Wind energy is the forerunner among the renewable energy sources, and by the end of 2020, wind energy accounted for roughly 8.4\% of the total electricity used in the United States \citep{eia}. In various decision making tasks in wind energy, wind power curve plays an important role.  A power curve is a function that maps the relationship of wind speed and other environmental variables to the wind power output. A quality estimation of power curve has crucial practical implication for decision-making in many aspects, including wind power prediction and turbine performance evaluation \citep{Ding2019}.

\begin{figure}
\centering
\includegraphics[width = 0.5\linewidth]{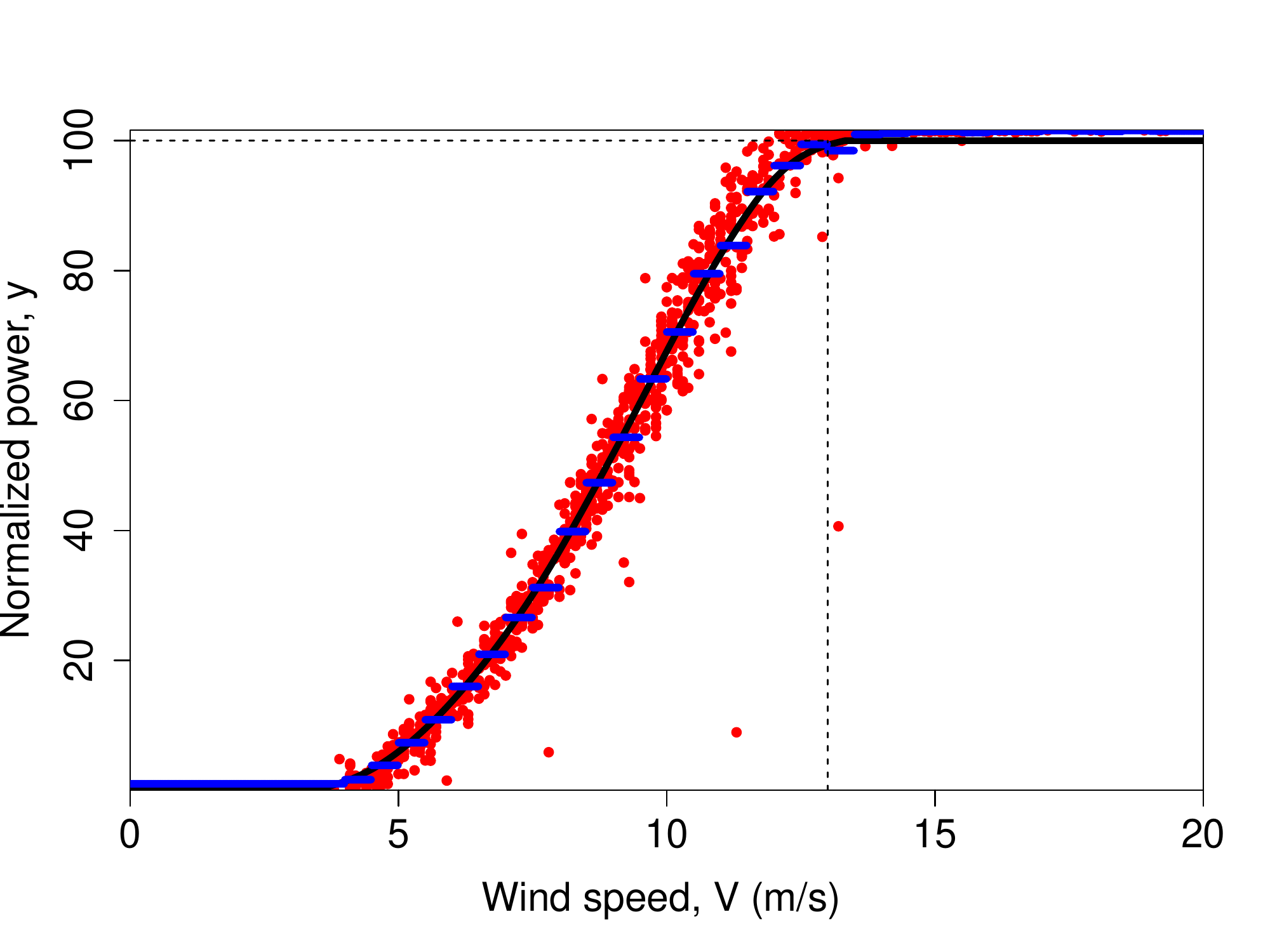}
\caption{A nominal wind power curve. Dots (red) denote the data; piecewise constant curve (blue) represents binning; smooth curve (black) is from smoothing on binning.} \label{Fig:powercurve}
\end{figure}

International Electrotechnical Commission \citep{IEC05} recommends a data-driven approach, known as the \emph{binning method}, to construct the power curve. The binning method considers the single input of wind speed, partitions wind speed into small bins, say, $0.5$ m/s, and uses the sample average of wind power data, whose corresponding wind speed falls into a bin, as the estimate of power response for that bin. The power curve in Figure \ref{Fig:powercurve} is generated using the binning method (with some simple post smoothing).

Research reported in \cite{bessa2012time} and \cite{Lee2015} identifies that wind power production is not limited to the effect of wind speed, but also depends on other factors such as wind direction, air density, etc. \cite{bessa2012time} and \cite{Lee2015} both used the kernel regression and kernel density estimation approaches. But \cite{bessa2012time} handle up to three input variables, while \cite{Lee2015} use a new additive-multiplicative model structure, referred to as AMK, that can in principle take in as many inputs as possible. The actual number of inputs used in \cite{Lee2015} is up to seven.  In Chapter 5 of \cite{Ding2019} and its companion \texttt{DSWE R package} \citep{DSWE-Package}, more nonparametric regression methods are provided, including those based on the smoothing splines (SSANOVA) \citep{Gu2013Book, Gu2014Package}, Bayesian trees (BART) \citep{Chipman2010}, k-nearest neighbors (kNN) \citep{hastie2009}, and support vector machines (SVM) \citep{Vapnik2000}.  Thus, the nature of power curve modeling falls squarely under the umbrella of nonparametric regression.

During our research, we encountered a problem in wind power curve modeling, which we explain through the following real-life example. We were given (by a wind company) a set of wind/weather inputs---wind speed, its standard deviation, wind direction, ambient temperature---denoted by $\bs x$, and the turbine power output data, denoted by $y$; all collected in 2015. We were asked to train a model and then predict $y$ using the $\bs x$ values collected in 2016 (the first six months).  The actual $y$ values of 2016 were withheld by the company for evaluation. The binning method is the company's practice and used as the benchmark. The company (in fact any company) would only consider adopting a new method, if the new method outperforms their current benchmark in testing.  We chose the AMK method mentioned above, as it was then demonstrated as the most competitive method. We use a five-fold randomized cross validation to train the AMK model, including both variable selection and parameter estimation.  Using a forward stepwise variable selection, all four aforementioned inputs are selected as important. The resulting AMK model has a root mean squared error (RMSE) 30\% smaller than the binning RMSE, based on the same five-fold randomized cross-validation using the 2015 data. However, when both models (binning and AMK) are applied to the 2016 data, the AMK model produces an RMSE which is 5\% higher than binning.  This is not a unique problem to AMK.  Should we use another nonparametric regression method included in the \texttt{DSWE R package}, a similar phenomenon will be observed, that is, all of them outperform binning, with a comfortable margin, when tested using 2015 data through a randomized cross validation, but all of them either fails to outperform binning or see their margin of improvement significantly diminished when tested using the 2016 data.

Exploring the literature \citep{roberts2017, meyer2018,sheridan2013}, we found that we are not alone---the problem encountered is a case of a common issue in nonparametric regression, known as \emph{temporal overfitting}~\citep{meyer2018}. This overfitting is caused due to the temporal autocorrelation in the data.  In the wind application, both $\bs x$ and $y$ are autocorrelated. Temporal overfitting differs from the usual notion of overfitting; the latter occurs when a model fits well to the training data, but performs worse on a random holdout data. The usual overfitting can generally be avoided using cross-validation, as cross-validation error tends to estimate the generalization error when the input and error processes are independent and identically distributed (i.i.d.); see \citet[Chapter 7]{hastie2009}. For temporal overfitting, however, the model generalizes well for the test datasets originating from the same time domain as the training dataset, that is, there is no temporal extrapolation when moving from training to test inputs and the data points in training and test sets are temporally close. But, the model's performance seriously deteriorates when the test dataset is from a different time domain (temporal extrapolation), precisely as we observed in the wind power curve modeling.

We can classify test sets and their corresponding test errors into two categories. For convenience, let us call the two test errors as \emph{in-temporal} and \emph{out-of-temporal} test errors, respectively, based on whether the test datasets arise from the same or a different time domain than that of the training dataset. In the earlier example, when a model is trained and tested on the 2015 data through a randomized cross validation, the corresponding test error is the in-temporal error, whereas when a model is trained using the 2015 data and tested using the 2016 data, the corresponding test error is the out-of-temporal error. To be clear, temporal overfitting is \emph{not} concerned with the training error like the usual overfitting problem. Rather, it is concerned with the aforementioned two types of \emph{test errors}.
	
In this work, we establish a Gaussian process (GP)-based method to deal with temporal overfitting. We split our functional model into two components: a time-invariant component and a time-varying component, each of which is modeled as a GP. The GP model in and of itself does not remove temporal overfitting; we make use of a subsampling scheme from the Bayesian statistics literature, known as thinning, for model inference that reduces the adverse impact of temporal overfitting. Thus, the main contributions of this work is to highlight the temporal overfitting problem in power curve modeling and propose a GP-based inference strategy to overcome the problem. Our numerical studies show that the proposed method performs significantly better for out-of-temporal predictions, as compared to the existing power curve models and other statistical approaches (those described in Section \ref{Section:ProbStatement}) that can be used to deal with the temporal overfitting problem. 

The rest of the paper is organized as follows. In Section \ref{Section:ProbStatement}, we describe the problem statement and the major schools of thought in dealing with temporal overfitting. Section \ref{Section:Method} presents our proposed method. Section \ref{Section:Application} provides the empirical evidence on performance of our method using wind turbine datasets. We conclude our research in Section \ref{Section:Disc}.
	
\section{Problem statement and relevant literature} \label{Section:ProbStatement}
We consider a nonparametric regression problem, where $Y_i \in \mathbb{R}$, $ X_i \in \mathbb{R}^d $, and $ u_i \in \mathbb{R}$, and
\begin{equation}\label{equ1}
	Y_i = f(X_i) + u_i,
\end{equation}
which has the following features:
\begin{enumerate}
	\item The form of $f(\cdot)$ is unknown and differs in applications, making nonparametric approaches more appropriate.  Furthermore, there are sufficient data pairs, $ \{y_i,x_i\}_{i=1}^N $, enabling the nonparametric treatment.
	\item The input is multivariate, but the number of input variables that are causal for the response is unknown.
	\item The input $X$ and the error $u$ can be considered physically independent with each other.
	\item The data are observational, not experimentally designed or controlled. They are sequentially collected from an ongoing physical process  over time. The index, $i$, corresponds to time. Both the input variables in $X$ and the error process, $u$, are temporally autocorrelated over the time index $i$.
\end{enumerate}

Our focus here is on \#4, because without the temporal autocorrelation, the problem falls under standard nonparametric regression.	

\begin{figure}
	\spacingset{1}
	\centering
	\includegraphics[width=\linewidth]{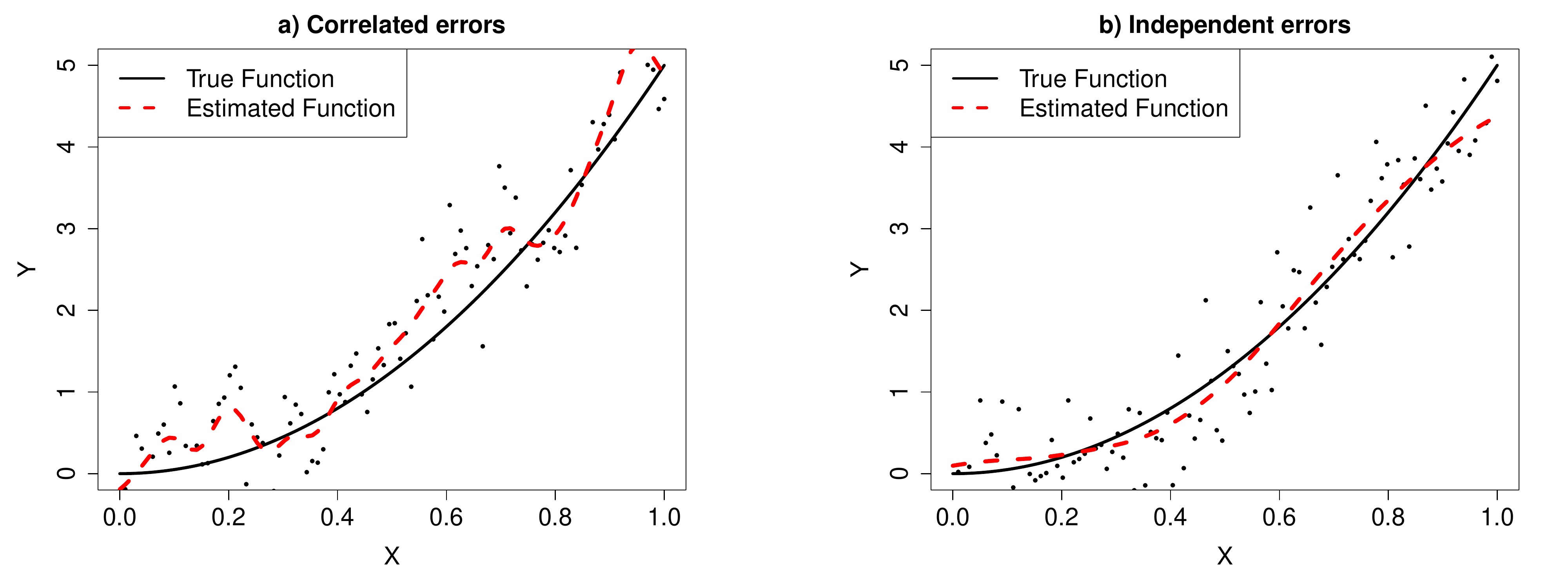}
	\caption{Effect of correlation between input variable and error on functional estimate:  a) correlated errors; b) independent errors. We use $f(x) = 5x^2$. The correlated error sequence is generated using a zero mean GP with input $x$ and an exponential kernel with a lengthscale of 0.05.} \label{Fig:tempoverfit}
\end{figure}

A natural question is how the temporal autocorrelation in the data causes the overfitting, even though time is not explicitly considered in the model (only implicitly tagged with $X_i$) and the function of interest is the relationship between the input variables $X$ and the response $Y$. 	To address this question, we consider a closely related problem in the statistics literature---when the error process is correlated with some input variables, namely $u = u(X)$.  If one applies the standard statistical learning techniques without accounting for the correlation between the error and the input variable, one may get an overfitted functional estimate as shown in the left panel of Figure \ref{Fig:tempoverfit}. The curve is fitted using a kernel regression with a direct plug-in (DPI) bandwidth estimate \citep{ruppert1995}. One can find similar plots in \cite{Opsomer2001}, \cite{debrabanter2011}.
	
The problem of temporal overfitting can be thought of a case when the errors are correlated with the input variables. We know that when two random variables change slowly over time, it could result in a spurious correlation among these two quantities even if they are independent, as in the case under our consideration; see \#3 and \#4 in the problem setup. This is to say, when the input variables and errors are autocorrelated in time, it would create a correlation among them, and may consequently create the overfitting effect as shown in Figure \ref{Fig:tempoverfit}. The presence of the nuisance input variables, which are not causal to the response, further aggravates the problem. An input variable could turn out to be `seemingly important,' owing to the temporal autocorrelation in both the response and the nuisance variable, even when it is not a causal variable. The more input variables considered for modeling, the more likely a non-causal variable to be selected because of the presence of temporal autocorrelation, leading to poor generalization when the test data points are from a different time domain \citep{roberts2017}.
	
There are various methods studied in the literature to handle the problem of autocorrelation in error and/or input; \cite{Opsomer2001} provides a survey of the methods up to two decades ago. One class of methods, developed specifically for kernel regression, are based on directly modifying the bandwidth estimation technique; see, for example, \cite{Altman1990} and \citet{debrabanter2011,debrabanter2018}. This is generally done by modifying the criterion for computing the optimal bandwidth such as asymptotic squared error (ASE). One of the criticisms for these kernel-based methods, as discussed by \cite{Opsomer2001}, is their inability to handle multivariate inputs. This limitation persists even in the recent literature \citep{debrabanter2018}, which still considers a univariate input while developing their bandwidth estimation. \citet{debrabanter2018} touch upon extensions to multivariate inputs only in their discussions section. Thus, this class of method is not directly applicable to our nonparametric regression problem where multiple input variables are dealt with.
		
There are two other schools of thought applicable to our problem setup. One is known as \emph{pre-whitening} \citep{xiao2003,geller2018} and the other is through some modification of the cross-validation error \citep{chu1991,burman1994,racine2000,Opsomer2001, rabinowicz2020}.

The idea of pre-whitening is to preprocess the response itself such that the resulting data has a white noise \citep{xiao2003,geller2018}. More specifically, pre-whitening is to model $ u_i $ using an invertible linear process,
\[u_i = \Sigma_{j=0}^{\infty}c_j\varepsilon_{i-j},\]
where $ \varepsilon $'s are white noises. Then, one can map $ u_i $'s to $ \varepsilon_i $'s using the inverse process. Practically, one would require to fit a regression model $ \hat{m}(X) $ and then compute the residuals $ \hat{u}_i = Y_i - \hat{m}(X_i)  $.  An autoregressive model of a suitable order can be estimated for $ \hat{u}_i $. Subtracting the estimated autoregressive part of $ \hat{u}_i $ from $ Y_i $, so as to remove the autocorrelation in $ Y $ due to $ u $, produces a modified response $ \bar{Y}_i $. Theoretically, this modified response $ \bar{Y}_i $ would be free from temporal autocorrelation and can be used as the response for the final model. One major challenge in this method is the presence of some nuisance variables in the data. According to \cite{roberts2017}, these nuisance (non-causal) variables can mask the autocorrelation in the residuals, that is, the temporal autocorrelation of the residuals gets modeled through the autocorrelation in the  nuisance variables. If that happens, one would underestimate the autocorrelation in the residuals, resulting in minimal or even no changes in the modified response  $ \bar{Y}_i $. As a result, one gets little or no improvement on the estimate of the regressor.

The idea of modifying cross-validation is arguably a more general framework to deal with correlated errors or temporal overfitting. This branch of methods is also known as $h$-blocking or $h\nu$-blocking and, more recently, as leave-time-out or time-split cross-validation. The idea is to do cross-validation on temporal blocks of data rather than random samples. \cite{chu1991,burman1994,racine2000} explore this idea under different settings. The time-blocked cross-validation idea is advocated by \cite{roberts2017} and adopted in \cite{meyer2018}.

Related to the idea of modifying cross-validation but unlike the previous works, \cite{rabinowicz2020} proposed a modification for the cross-validation error by adding a correction factor to account for the correlation. They motivate the problem using a linear mixed effect model where some of the effects stay the same, whereas other effects change from training to test datasets. The model formulation in \cite{rabinowicz2020} bears certain similarity to ours, as we split the regression function into a time-invariant and time-dependent component (to be presented in the next section). A key difference is that the inputs to the time-invariant component in our model are autocorrelated, while the inputs to \cite{rabinowicz2020}'s fixed effect term are i.i.d. The rest of the treatment in our method also differs substantially from that in \cite{rabinowicz2020}. For instance, our inference method does not require cross-validation.

In the case study, we compare our proposed method with the pre-whitening method, the time-split cross-validation, and \cite{rabinowicz2020}'s method. We find that our method consistently outperforms these available approaches when testing on data that are outside the time domain covered by the training data.
	
\section{Proposed method}\label{Section:Method}
 In this section, we describe our proposed method to mitigate the problem of temporal overfitting while fitting a nonparametric regression model.
	
\subsection{The model}
Given a dataset $\mc{D} = \{y_i,\bs{x}_i,t_i\}_{i=1}^N $, we consider the following model:
\begin{equation}\label{Eqn:Model}
y_i = f(\bs{x}_i) + g(t_i) + \epsilon_i,
\end{equation}
where for our target wind application, $ y $ is the power output of a wind turbine, $ \bs{x} $ is the $ d $-dimensional vector of environmental input variables, $ t $ denotes time, and $ \epsilon $ is i.i.d. Gaussian noise with zero mean and variance $ \sigma^2_\epsilon < \infty$. We deem that $f(\cdot)$ is a time-invariant function of  the input variables $ \bs{x}$ and $g(\cdot)$ is a temporally autocorrelated stationary stochastic process that contains the autocorrelated part of the residual. We stress that while $f(\cdot)$ is time invariant, $\bs{x}_i$ is time varying and autocorrelated.  

Recall that the motivation for this paper is to avoid temporal overfitting and improve the prediction accuracy under temporal extrapolation, that is, for out-of-temporal test sets as defined in Section \ref{Section:Intro}. In Equation \eqref{Eqn:Model}, $ f(\cdot) $ can explain the variance in the data that is carried over to a different time domain, as we assume that this function does not directly depend on time but only through the input variables. The variance which does not carry over to a different time horizon is modeled through the time-dependent term, $ g(\cdot) $. The rest is just i.i.d. noise. Thus, for out-of-temporal predictions, accurately identifying $f(\cdot)$ plays a key role in improving the accuracy.
	
Before providing further modeling details, we would like to elaborate on this model setup.  Among the three main approaches reviewed in Section~\ref{Section:ProbStatement}, the pre-whitening approach and the time-split cross-validation do not invoke a model like in Equation~\eqref{Eqn:Model}; rather they work directly with the model in Equation~\eqref{equ1}.  This is especially obvious in the pre-whitening approach, which is to whiten the autocorrelated $u$ and use that as the main apparatus to deal with temporal overfitting.  But \cite{rabinowicz2020}'s method does invoke a model of certain similarity to Equation~\eqref{Eqn:Model}. Specifically, \cite{rabinowicz2020} consider a generalized linear mixed model (GLMM) of the form,
\begin{equation}\label{Eqn:GLMM}
\boldsymbol{y} = \Phi\boldsymbol{\beta}+Z\boldsymbol{s} +\boldsymbol{\epsilon},
\end{equation}
where $\Phi$ contains the fixed effect covariates, meaning that the input $\boldsymbol x$ would be included through $\Phi$, and $Z$ contains the random effect covariates. The realization of the random effect $ \bs{s} $ can change from training to test cases.

As we pointed out earlier,  \cite{rabinowicz2020} assume the input variables in their fixed effect term, $ \Phi $, to be i.i.d. samples. In our setting, the input variables in $ \bs{x}_i $ are autocorrelated in time. In other words, the autocorrelation in $y_i$ in our process comes from two sources---the autocorrelation in both $ \bs{x}_i $ and $g(t_i)$.  We believe this difference is important and helps explain the difference in the outcome of applying both methods.  We revisit this point in Section \ref{Section:CVc} after presenting the numerical results.

Continuing with our modeling process, we model both $f(\cdot)$ and $g(\cdot)$ as realizations of stationary Gaussian processes (GPs) \citep{Rasmussen2006}. For $f(\cdot)$, it is assumed to be a sample from a GP with a mean function $ \mu(\cdot) $ and a covariance function $ k(\cdot,\cdot) $. Specifically, we use a constant mean function and a Mat\'{e}rn covariance function with a smoothness parameter of $\nu=1.5$ as follows:
\begin{equation}\label{Eqn:fx_prior}
\mu(\bs{x}) = \beta, \qquad k(\bs{x},\bs{x}') = \sigma_f^2 \Big(1+\sqrt{3}r\Big)\exp{\Big(-\sqrt{3}r\Big)}, \quad r = \sqrt{\sum_{\ell=1}^{d}{\frac{((\bs{x})_\ell - (\bs{x}')_\ell)^2}{(\bs{\theta})_\ell^2}}}
\end{equation}
where $ \beta $ is an unknown constant, $ \sigma_f^2 $ is the variance of $ f(\cdot) $, and $ (\cdot)_\ell $ denotes $ \ell^{th} $ component of a vector; for instance, $ (\bs{\theta})_\ell $ is the lengthscale parameter for the $ \ell^{th} $ covariate. We come to this specific choice by experimenting with four covariance functions---squared exponential, two Mat\'{e}rn kernels (smoothness of 1.5 and 2.5), and exponential kernel---and chose the best performing one, which is Mat\'{e}rn with $\nu=1.5$.  While there are some difference in performance, the differences are not that striking; see Supplementary Material S1.
	
The temporal part $ g(\cdot)$ is assumed to be zero mean with a covariance function denoted by $ q(\cdot,\cdot) $. We again use a Mat\'{e}rn covariance function with $\nu=1.5$ for $g(\cdot)$ given as follows:
\begin{equation}\label{Eqn:gt_prior}
	q(t,t') = \sigma_g^2\Big(1+\sqrt{3}\frac{\lvert t-t'\rvert}{\phi}\Big) \exp{\Big(-\sqrt{3}\frac{\lvert t-t'\rvert}{\phi}\Big)},
\end{equation}
where $ \sigma_g^2 $ is the variance of $ g(\cdot) $ and $ \phi $ is the lengthscale for time.
	
\subsection{Inference procedure}\label{Section:Inference}
The key is how to effectively estimate the two components in Equation~\eqref{Eqn:Model}. Recall that our data are observational, not from designed experiments in which the confounding effects can be distinguished through a careful selection of factor settings. Using the observational data, if one conducts the maximum likelihood estimation of the hyperparameters, $ (\beta, \sigma_f, \bs{\theta}, \sigma_g, \phi, \sigma_\epsilon )$, in Equation \eqref{Eqn:Model}, one would run into an identifiability issue---while attempting to learn the hyperparameters for both $f(\cdot) $ and $g(\cdot)$ together, it is difficult to tell whether the variance in the data is due to some input variables or due to time. We provide numerical evidence on worse performance of joint (direct) estimation in Section \ref{Section:DirectEst}.

Another possible explanation for the inferior performance of direct estimation comes from an intrinsic problem of GP regression. For simplicity, consider the model $y_i=\mu_0+f_0(\bs{x}_i)+g_0(t_i)+\epsilon_i$, where $\mu_0$ stands for the global mean value, and $f_0$ and $g_0$ are zero-mean GPs. Then there is a known identifiability issue for estimating $\mu_0$; see \cite{tuo2016theoretical} and Theorem 3 of \cite{wang2020prediction}. This issue in estimation does not affect the prediction properties for applications in which people combine all additive terms for making prediction, because this bias can be compensated by another bias in estimating $f_0$ and $g_0$; see \cite{tuo2018prediction} and Theorem 2 of \cite{wang2020prediction}. However, in the current context, the estimate of $g_0$ itself is critical for out-of-temporal predictions, because an inaccurate estimate of $g_0$ means that $\mu_0+f_0$ is also inaccurate, resulting in worse out-of-temporal predictions.

In order to overcome this problem, we decompose the model in Equation \eqref{Eqn:Model} as follows:
\begin{eqnarray*}
y_i &=& f(\bs{x}_i) + u_i , \\
u_i &=& g(t_i) +\epsilon_i .
\end{eqnarray*}
In the first step, we only focus on estimating  $ f(\cdot) $ such that the variation in $ y $ due to temporal correlation of $u$ does not get modeled. The residual left after subtracting the estimate of $ f(\cdot) $ would be used to estimate $ g(\cdot) $ and $ \sigma_\epsilon^2 $.

The first step is equivalent to learning a function with autocorrelated errors. To this end, we adopt an idea frequently used to reduce the autocorrelation in Markov chain Monte Carlo (MCMC) sampling schemes. The method is a subsampling scheme known as \emph{thinning}. The method retains one sample after every $ T $ time steps and discards all the samples in between to reduce the autocorrelation; hence the name, thinning, as this results in a thinned dataset. The number $ T $ is often referred as the thinning number.

Thinning retains only $1/T$ fraction of the original dataset and discards the rest. We would like to retain all the samples because each sample carries information about the function, which may not be otherwise available in other data points. So, instead of discarding the data, we put the training samples in $ T $ number of thinned data bins. Let us denote a data point  $ (y_i, \bs{x}_i, t_i) $ as $ \mc{D}_i $. Then, the first bin will have following data points, $\mc{B}_1 :=  \{\mc{D}_1, \mc{D}_{T+1}, \mc{D}_{2T+1}, \dots , \mc{D}_{\lfloor \frac{N}{T}  \rfloor T + 1}\}$, where $ \lfloor a \rfloor $ rounds $a$ down to its nearest integer. Generally, the $ j^{th} $ bin has the following data points:
\[\mc{B}_j :=  \{\mc{D}_j, \mc{D}_{T+j}, \mc{D}_{2T+j}, \dots , \mc{D}_{\lfloor \frac{N}{T}  \rfloor T + j}\}. \]
If $\lfloor \frac{N}{T}  \rfloor T + j > N$, then $(\lfloor \frac{N}{T}  \rfloor - 1) T + j $ would be last element for $ \mc{B}_j $. We have $T$ bins overall.

Thinning creates a temporal gap between two consecutive data points in a bin, and thus reduces the intra-bin temporal autocorrelation among the training points in any given bin. Hence, for the data points in the same bin, we could assume $ u $ to be independent Gaussian noise with some variance $ \sigma_{u}^2 < \infty $. Then, we can proceed to estimate $f$ using a likelihood function of the thinned data.
	
Let us denote the number of data points in $ \mc{B}_j $ by $ n_j $ and let $ \pi_j(\cdot) $ be the function that maps the index of the elements in set $ \mc{B}_j$  to the index of the original dataset $ \mc{D} $. Denote the response vector for $ \mc{B}_j $ by  $ \bs{y}^{(j)} = (y_{\pi_j(1)}, y_{\pi_j(2)},\dots,y_{\pi_j(n_j)})^\top  $. Let $ \mb{K}^{(j)} $ be the covariance matrix for $ \mc{B}_j $ such that the element in $ r^{th} $ row and $ s^{th} $ column is given by $  (\mb{K}^{(j)})_{rs} = k(\bs{x}_{\pi_j(r)},\bs{x}_{\pi_j(s)}) \ | \ r, s = 1\dots n_j$, where $ k(\cdot,\cdot) $ is the same as defined in Equation \eqref{Eqn:fx_prior}. Let $\mb{I} $ be an identity matrix of a proper size and $ \mathds{1} $ be a vector of ones of a compatible size. Then, the likelihood function for the $ j^{th} $ bin $ \mc{B}_j $ is given as follows:
\begin{equation}\label{Eqn:ll_bin}
\mc{L}_j= \frac{1}{\sqrt{(2\pi)^{n_j}\lvert \mb{K}^{(j)}+ \sigma_{u}^2 \mb{I} \rvert}} \exp{(-\frac{1}{2}(\bs{y}^{(j)}-\beta \mathds{1})^\top [\mb{K}^{(j)}+ \sigma_{u}^2\mb{I}]^{-1} (\bs{y}^{(j)}-\beta \mathds{1}))} \ .
\end{equation}

One could use any individual bin and its likelihood function to estimate the hyperparameters for function $f$, but doing so makes use of only a fraction of the original data, which is not ideal.  Intending to make full use of all the data for estimating $f$, our approach is to create a pseudo-likelihood function as the product of the likelihood functions of individual bins, namely $\prod_{j=1}^{T}\mc{L}_j$.  As such, the hyperparameters of function $f$ can be estimated by maximizing this pseudo-likelihood function, as follows:
\begin{equation}\label{Eqn:ll_product}
(\hat{\beta},\hat{\sigma}_f^2,\hat{\bs{\theta}},\hat{\sigma}_{u}^2) = \arg\max \prod_{j=1}^{T}\mc{L}_j \ .
\end{equation}
The temporal correlation between different bins, namely the inter-bin temporal correlation, will still exist, because temporally neighboring data points are now in different bins. But the construction of the pseudo-likelihood function ignores the inter-bin correlations. In other words, from the lens of this pseudo-likelihood function, the bins are not related at all. Optimizing this pseudo-likelihood function forces the estimation of $f$ to the thinned data and will not be affected by the temporal autocorrelation.

The hyperparameters in Equation \eqref{Eqn:ll_product} can be estimated using any optimization routine. In practice, we work with the logarithm of the likelihood function, which changes the finite product structure of the likelihood function to a finite sum. Finite sum functions can also be optimized using parallel processing, that is, each of the summand function can be evaluated independently on a different computing core, which could reduce the computation time.
	
The hyperparameters therefore estimated will not reflect or minimally reflect the temporal correlation due to $ u $. Once the hyperparameters are estimated, we combine the bins back to create a single dataset and use the single dataset for predictions. We use the following notation: $ \mb{K} $ is the covariance matrix for all the training points with its element in the $ i^{th} $ row and $ j^{th} $ column given as $ (\mb{K})_{ij} = k(\bs{x}_i,\bs{x}_j) \ \mid i,j = 1\dots N$, $ \bs{r}(\bs{x})  = (k(\bs{x},\bs{x}_1), \dots ,k(\bs{x},\bs{x}_N))^\top $ is the correlation vector between any point $ \bs{x} $ and all the training points, and $ \bs{y} = (y_1, \dots ,y_N)^\top $ is the vector of response for all the training points. The value of $ \hat{f}(\cdot) $ can be calculated at a given point $ \bs{x} $ as follows:
\begin{equation}\label{Eqn:fpred}
\hat{f}(\bs{x}) = \hat{\beta} + \bs{r}^\top(\bs{x}) [\mb{K} + \hat{\sigma}_{u}^2\mb{I}]^{-1}(\bs{y} - \hat{\beta} \mathds{1}).
\end{equation}

\subsection{Estimating $g(t)$}
	
Let $ \bs{\hat{f}} = (\hat{f}(\bs{x}_1),\dots,\hat{f}(\bs{x}_N))^\top $ be the vector of the estimate of $ f(\cdot) $ for all the training points. The vector of residual $ \bs{\hat{e}} $ is calculated as: $\bs{\hat{e}} = \bs{y} - \bs{\hat{f}} $. Each of the residual is associated with a time point. Denote by $ \hat{e}_t $ the residual for time point $ t $. We use these residuals as the response for estimating $ g(t) $ and $ \sigma^2_\epsilon $.
	
Here we assume that the autocorrelation in time decays much faster as compared with the overall time span of the training dataset.  This is definitely reasonable for our target wind applications, because the autocorrelation in wind speed or other environmental variables only persists in the order of hours~\citep[Figure 2.3]{Ding2019}, while our training data spans from a number of months to more than a whole year. For this reason, we do not need all the training points to compute a global estimate for $ g(\cdot) $. Instead, we compute a local estimate of $ g(t^*) $ at $ t^* $, based only on the training points in the neighborhood of $ t^* $. Doing this substantially reduces the computational burden for estimating $ g(\cdot) $.
	
We use a neighborhood based on the thinning number $ T $, as temporal autocorrelation would be small after a lag of $ T $ time units. Thus, we only include the training points that are within $ \pm T $ time units from $ t^* $, while estimating $ g(t^*) $. Moreover, there is no need to estimate $ g(t^*) $ if there happens to be no training points in the  $ T$-neighborhood of $ t^* $.
	
Let us define an index set for training points close to point $ t^* $ as $ J^* = \{  j : \  \lvert t^* - t_j \rvert \leq T; j = 1\dots N  \} $. Denote by $ \mb{Q}^* $ the covariance matrix formed by the time points in $ J^* $ such that $ (\mb{Q}^*)_{ij} = q(t_i,t_j); i,j \in J^* $. Here $ q(\cdot,\cdot) $ is the same as defined in Equation \eqref{Eqn:gt_prior}. Also, denote by $ \hat{\bs{e}}^* $ a  vector of residuals with its $ j^{th} $ component $(\hat{\bs{e}}^*)_j = e_{t_j} ; j \in J^*  $. The hyperparameters for $ q(\cdot,\cdot) $ and $ \sigma_\epsilon^2 $ is estimated based on the value of $ t^* $, and is given as
\begin{equation}\label{Eqn:ll_gt}
	(\hat{\sigma}_g^{2},\hat{\phi},\hat{\sigma}_{\epsilon}^2) =  \arg\max \frac{1}{\sqrt{(2\pi)^{(\lvert J^* \rvert)}\lvert \mb{Q}^{*}+ \sigma_{\epsilon}^2 \mb{I} \rvert}} \exp{(-\frac{1}{2}\bs{\hat{e}}^{*^\top} [\mb{Q}^{*}+ \sigma_{\epsilon}^2\mb{I}]^{-1} \bs{\hat{e}}^{*})}.
\end{equation}
Let $ \bs{s}^* $ denote a covariance vector such that its $ j^{th} $ component is $ (\bs{s}^*)_j = q(t^*,t_j); j \in J^* $. Once the hyperparameters are estimated, $\hat{g}(t^*)$ is given as follows:
\begin{equation}\label{Eqn:gpred}
	\hat{g}(t^*) = {\bs{s}^*}^\top [\mb{Q}^* + \hat{\sigma}_\epsilon^2 \mb{I}]^{-1} \hat{\bs{e}}^*
\end{equation}
\subsection{Predictions}
Once $\hat{f}(\cdot)$ and $\hat{g}(\cdot)$ are estimated, they do not have to be used together in a prediction.  If one wants to predict at a time $t^*$, which is in the far distant future and temporally far away from any training data points, then one only needs $ \hat{f}(\bs{x}^*)$, as $\hat{g}(t^*) $ is going to be zero.  The condition to decide whether $t^*$ is temporally enough far away from the training data is to check whether there exists a training data point, $t_i$, such that $ \lvert t^* - t_i \rvert \leq T $.  If no, then $t^*$ is temporally far away. We refer to this type of prediction under temporal extrapolation as \emph{out-of-temporal prediction}. By contrast, the\emph{ in-temporal prediction} refers to predictions over a time $t^*$ not temporally far away from the training data points.

Of course, a user does not have to check this condition.  One can just use $ \hat{f}(\bs{x}^*) + \hat{g}(t^*) $ to predict at any test point $ (\bs{x}^*, t^*) $, regardless of where $t^*$ is.  The $\hat{g}(\cdot)$ term takes the temporal distance between $t^*$ and the training data points into consideration and will reduce to zero when $t^*$ is temporally distant from the training data.  Simply put, our model can adapt for out-of-temporal versus in-temporal prediction without a user's active involvement.
	
\subsection{Choice of thinning number}
The choice of thinning number $ T $ is important to reduce the temporal autocorrelation within a data bin. If $ T $ is very small, we may still have overfitting problem. On the other hand, if $ T $ is very large, the number of data points in each bin may be too low to learn $f(\cdot)$ accurately. The choice of $ T $ needs to provide a trade-off between these two aspects. Since, we want to ensure that the temporal autocorrelation is sufficiently reduced, we choose the thinning number as the smallest lag such that the absolute value of the partial autocorrelation function (PACF) for each of the covariates is less than two standard errors of the PACF for a sequence of $N$ i.i.d Gaussian noise, which is considered to be statistically insignificant at the 95\% significance level. In other words, the value of $ T $ is  given as follows:
\begin{equation}\label{Eqn:Thinning}
T = \max_{\ell = 1,\dots,d} \min_{h} (\  \mid\text{PACF}_{(\boldsymbol{x})_\ell}(h)\mid \ \leq 2/\sqrt{N} ),
\end{equation}
where $\text{PACF}_{(\boldsymbol{x})_\ell}(h)$ is the PACF for covariate $\ell= 1, \dots, d $ for lag $h$. We tested the choice of thinning number in Section \ref{Section:ThinExp} through a sensitivity analysis on real datasets. 

\section{Case study: Application to wind turbine datasets}\label{Section:Application}
We present two case studies for modeling wind power curves to validate the performance of the proposed method. Both datasets are publicly available. The first case study is based on four datasets. We refer to the first case study as \emph{Case Study I}. The second case study is on a larger number of datasets (thirty turbines) but the data available are for a shorter period of time. We refer to the second case study as \emph{Case Study II}.
	
\subsection{Datasets}\label{Section:Datasets}
Case Study I uses four datasets available on the book website of \cite{Ding2019} (\url{https://aml.engr.tamu.edu/book-dswe/dswe-datasets/}, Dataset 6).  They are associated with four turbines, denoted as WT1 to WT4. The first two datasets are from inland and the remaining are from offshore wind turbines. Each turbine has four years of data collected at a 10-minute frequency. The data for the inland turbines (WT1 and WT2) span from 2008 to 2011, and those for the offshore turbines (WT3 and WT4) extend from 2007 to 2010.  Each dataset has five environmental input variables along with the response ($ y $) and time stamp ($ t $) for each data point. The inland turbines have the same input variables: wind speed ($ V $), wind direction ($ D $), air density ($ \rho $), turbulence intensity ($ I $), and wind shear ($ S $). The offshore turbine datasets contain humidity ($ H $) instead of wind shear and have the rest of the four variables same as the inland turbines. One can easily see that all these variables are temporally autocorrelated by plotting their partial autocorrelation function (PACF) plots. One such example for WT1 is provided in the Supplementary Material S2.

Each of the datasets has missing data points. The exact number of data points is given in Table \ref{dataSpecs}. The number of data points are about 50\% when compared to the scenario where turbines produce power at all the times (for every 10 minute interval, we have a positive power), which is an ideal condition and not observed in practice.
There are two major causes of missing data points: 1) wind speed is either below cut-in speed or above cut-out speed, so there is no power production; 2) wind conditions are favorable but the power output is either curtailed or zero because of grid commitments or operational issues.

\begin{table}
\spacingset{1}
\centering
\caption{Description of the main study datasets.}\label{dataSpecs}
\resizebox{\textwidth}{!}{
\scriptsize
\begin{tabular}{c | c c c c}
\hline
\textbf{Dataset} & WT1 & WT2 & WT3 & WT4\\
\hline
Type of  wind turbine & Inland & Inland & Offshore & Offshore \\ 
Time period & 2008--2011 & 2008--2011 & 2007--2010 & 2007--2010 \\ 
Covariates  &  $ \{V,D,\rho,I,S\} $  &  $ \{V,D,\rho,I,S\} $  &  $ \{V,D,\rho,I,H\} $  &  $ \{V,D,\rho,I,H\} $ \\ 
Number of data points & 96,824 & 89,730 & 113,378 & 110,556 \\
\hline		
\end{tabular}
}
\end{table}

Case Study II is based on thirty wind turbines. These datasets are available at \url{https://github.com/TAMU-AML/Datasets/tree/master/TemporalOverfitting}. The input variables are the same as that of the inland turbines in Case Study I. These thirty datasets can be further classified into groups of ten, as the ten turbines in the same group share the same meteorological tower. The meteorological towers measure wind speed at multiple heights, wind direction, ambient pressure and temperature. The multi-height wind speed measurements are used to calculate the wind shear. Ambient pressure and temperature are used to calculate the air density. Wind direction data are also taken from the meteorological towers. Each of the turbines also measure the wind speed at their nacelle. The data for wind speed and power are collected at individual turbines, and turbine's wind speed data is used to calculate the turbulence intensity. The first meteorological tower has a slightly longer duration of data than the other two towers; see Table \ref{Table:Data_Duration}.
	
For both case studies, we divide  the  datasets into three temporally disjoint datasets: $ \mc{T}_1, \mc{T}_2, \text{ and }  \mc{T}_3 $.  In Case Study I, $ \mc{T}_1 $ corresponds to the first two years, and $\mc{T}_2 \text{ and }  \mc{T}_3 $ has the data for the third and fourth year, respectively. For example, for the inland turbines (WT1 and WT2), $ \mc{T}_1$ contains the data for the years 2008 and 2009, and $ \mc{T}_2 \text{ and }  \mc{T}_3$ correspond to the year 2010 and 2011, respectively. In Case Study II, the duration corresponding to $ \mc{T}_1, \mc{T}_2, \text{ and }  \mc{T}_3 $ are listed in Table \ref{Table:Data_Duration}. We select a few methods including our proposed method to learn the power curve for each wind turbine using their corresponding $ \mc{T}_1 $ datasets. The learned model is then used to do out-of-temporal predictions on $ \mc{T}_2 \text{ and }  \mc{T}_3$.
	
\begin{table}[h]
\spacingset{1}
\centering
\caption{Duration of the datasets and temporal partitions for Case Study II.}\label{Table:Data_Duration}
\resizebox{\textwidth}{!}{
\begin{tabular}{c | c | c | c | c}
\hline
\textbf{Turbines} & Total duration & $ \mc{T}_1$ duration 	& $ \mc{T}_2$ duration 	& $ \mc{T}_3$ duration 	\\
\hline
1 to 10 & Apr 29, 2010--Oct 31, 2011 & Apr 29, 2010--Nov 30, 2010 & Dec 1, 2010--May 31, 2011 & Jun 1, 2011--Oct 31, 2011 \\
\hline
11 to 20 & Jul 30, 2010--Oct 23, 2011 & Jul 30, 2010--Dec 31, 2010 & Jan 1, 2010--May 31, 2011 & Jun 1, 2011--Oct 23, 2011 \\
\hline
21 to 30 & Jul 30, 2010--Oct 20, 2011 & Jul 30, 2010--Jan 31, 2011 & Feb 1, 2011--May 31, 2011 & Jun 1, 2011--Oct 20, 2011 \\
\hline
\end{tabular}
}
\end{table}
	
\subsection{Implementation and comparison}\label{Section:Implementation}

For implementing our method, we use all the five input variables in the model as the starting point but subset selection is part of the learning task; see Section \ref{Section:Datasets} for description of the input variables. Also, since wind direction is a circular variable, we embed it into a two-dimensional Euclidean space by using its \emph{sine} and \emph{cosine} transformations. We standardized all the input variables by subtracting their respective sample mean and dividing them by their sample standard deviation. Standardization of the inputs ensure that the importance of the variables would be clear from their respective lengthscale. A large lengthscale would imply that the input variable is not important in predicting the response.

We proceed by computing the thinning number $ T $ for each of the datasets as per Equation \eqref{Eqn:Thinning}. The computed value of $ T $ for the four datasets in Case Study I is shown in Table \ref{Table:thinning}. Taking the thinning number for WT1 as an example, the value of 12 would be equivalent to 2 hours, as the data points are collected every 10 minutes. This implies that two data points with less than 2 hours of time gap would not be kept in the same bin. Since there are missing data points in the dataset, the time of 2 hours serves as the minimum time gap between two intra-bin points. The actual temporal gap for some of the consecutive data points in a bin would be higher. We bin the datasets using their respective thinning number given in Table \ref{Table:thinning}, and use Equations \eqref{Eqn:ll_bin} and \eqref{Eqn:ll_product} to estimate the hyperparameters for the time-invariant function $ f(\cdot) $, as described in the previous section. Once we have the hyperparameter estimates for $ f(\cdot) $, we compute the out-of-temporal predictions on $ \mc{T}_2 $ and $ \mc{T}_3 $ using just $ f(\cdot) $, as given in Equation \eqref{Eqn:fpred}.
\begin{table}
\spacingset{1}
\centering
\caption{Thinning number, $ T $, for each of the four datasets.} \label{Table:thinning}	
\begin{tabular}{c | c c c c}
\hline
Dataset & WT1 & WT2 & WT3 & WT4 \\
\hline
Thinning Number & 12 & 14 & 22 & 22 \\
\hline		
\end{tabular}
\end{table}

We compare our proposed method with two categories of methods: the first category is for the nonparametric power curve methods that do not consider the issues of temporal overfitting, and the second category is for the approaches that address the serial autocorrelation and temporal overfitting issues, which are reviewed in Section \ref{Section:ProbStatement}.

In the first category, we use the following three methods---the IEC binning method, k-nearest neighbors (kNN), and additive multivariate kernel (AMK) by \cite{Lee2015}, but with the focus on out-of-temporal predictions only. We include the binning method because it is the industry baseline. If a proposed method cannot outperform this baseline, the value of that method will be called into question. kNN and AMK are included because the comparison presented in Chapter 5 of \cite{Ding2019} shows that these two methods do a better job than other nonparametric regression methods.

In the second category, we again consider three methods. The first is the time-split cross-validation \citep{burman1994,racine2000,roberts2017}, the second is based on \cite{rabinowicz2020}'s corrected cross-validation error, and third one is based on pre-whitening of the response \citep{xiao2003,geller2018}.

The implementation of the binning method is straightforward; we use a 0.5 m/s bin-width (the IEC standard). For kNN and AMK, one would have to do variable selection. We employ a forward stepwise subset selection using a five-fold cross validation to get the best subset of input variables. The AMK method by \cite{Lee2015}, which was specifically developed to model the wind turbine power curves, uses a kernel regression method. In their paper, \cite{Lee2015} consider additive combinations of trivariate kernels, keeping the first two variables common in all the additive terms and varying the third variable. They kept the common variables fixed as wind speed and wind direction. We modify their method by only fixing the first variable as wind speed and let the data decide the second common variable for the additive terms, while still using trivariate kernels. The analyses for both kNN and AMK are done in \texttt{R} using the \texttt{DSWE package} \citep{DSWE-Package}.

\begin{figure}[h]
\spacingset{1}
\centering
\includegraphics[width = 0.75\linewidth]{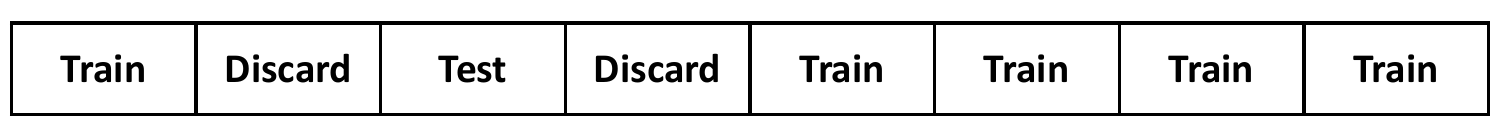}
\caption{A schematic of time-split cross-validation. Each block represents a group of temporally adjacent data points.} \label{Fig:timesplitvalidation}
\end{figure}

In order to do time-split cross-validation, we use kNN as the base method, but modify the cross-validation scheme. We divide the temporally ordered data into small blocks of size $ T $---same as the thinning number used in our proposed method. Instead of randomly sampling training and test datasets, we select training and test samples from the temporal blocks in a way such that if a particular temporal block is in test set, its neighboring blocks must not be in the training set. This splitting ensures that there is low temporal correlation between training and test datasets. A schematic of time-split cross-validation is given in Figure \ref{Fig:timesplitvalidation}. We do a five-fold time-split validation by sampling different test datasets and refer to the resulting method as TS-kNN, namely time-split kNN.

Time-split cross-validation can also be clubbed with any other base method such as AMK; however, given the size of the datasets, doing so is computationally expensive. For example, it took more than 10 hours to find the best subset of variables and compute the optimal bandwidths for WT1 when AMK method was clubbed with time-split cross-validation whereas TS-kNN took less than 15 minutes to do the same. Also, the result obtained using time-split validation on AMK did not show any significant improvement over standard AMK which uses a direct plug-in (DPI) approach by \cite{ruppert1995} for estimating the bandwidths. So, we decide to proceed with only kNN as the base method.

For \cite{rabinowicz2020}'s method, we again use kNN as the base method for the same reasons described for time-split validation; cross-validation with AMK is computationally prohibitive for the data size at hand. We refer to the resulting kNN method as CVc-kNN. The CVc method relies on estimating the covariance matrix of the response, $ Cov(\bs{y},\bs{y}) $, conditioned on the input variables. Here we run into a problem. For our problem setting, we need to estimate the covariance in $ \bs{y}$ due to $g(t)$ but the covariance in the response data are caused by the autocorrelation in both $ \bs{x} $ and $g(t)$. To apply the idea of CVc, we come up with an \emph{an hoc} procedure, which is to first fit a one-dimensional kNN model to the response data and then use the residual to compute $ Cov(\bs{y},\bs{y}) $. The thought behind is that the single input of wind speed is the most important variable in power curve models. Subtracting its effect would remove a major portion of the covariance in $ \bs{y} $ due to the temporal autocorrelation in $ \bs{x} $. Once $ Cov(\bs{y},\bs{y}) $ is estimated, we keep it fixed and do a forward subset selection, based on CVc, to find the best variables subset and corresponding hyperparameter ($ k $). It is apparent that the lack of a quality estimate of $Cov(\bs{y},\bs{y}) $ under our problem setting presents a major roadblock to the effective application of CVc (more discussion in Section~\ref{Section:CVc}).

The last method is based on pre-whitening of the response \citep{xiao2003,geller2018}. We follow the steps in \cite{xiao2003}, but with AMK as our base method. Here, we use AMK not only because AMK as the base method is a better choice than kNN, but also because \cite{xiao2003} uses a kernel-based local polynomial method, meaning that AMK is more compatible. We did not use AMK to be clubbed with time-split cross-validation, due to heavy computation that would have otherwise resulted. But for pre-whitening, as it does not require cross-validation to compute the optimal bandwidths, this computational burden is not there.  In other words, AMK can be used along with DPI approach once the response is modified.

Specifically, we fit an AMK model to the datasets and estimate the residuals at the training points. We use these residuals to fit an autoregressive (AR) model of order $ T $---the computed thinning number. We then modify the response by subtracting the estimated autoregressive component of residuals from the response. The modified response is then used to build a new AMK model, which would be used for predictions. Since we have already obtained the best input variable subset using forward subset selection for AMK, we use the same subset of input variables for the model and do not carry out subset selection again. We refer to this pre-whitened AMK model as PW-AMK.

\subsection{Results for Case Study I}\label{Section:Results1}
We present the results in two parts, corresponding to the two categories of methods. The first part compares our method with the power curve methods in Category 1, in which we use the binning method as the benchmark and compute performance improvement over binning for each method. The performance criteria is the root mean square error (RMSE) on a test dataset. Our proposed method is referred to as tempGP in the comparison.

\begin{table}[h]
\spacingset{1}
\caption{A comparison table for out-of-temporal RMSE for dataset $ \mc{T}_2 $ } \label{Table:RMSE_T2}
\centering
\begin{tabular}{c | c c c c c}
\hline
\textbf{Dataset} & Binning &  kNN &  AMK &    tempGP ($ f(\bs{x}) $)\\
\hline
\multirow{2}{2.5em}{WT1}   &  \multirow{2}{2.5em}{4.98}  & 4.96 &  4.35& \textbf{3.52}  \\
& & (0.4\%) &  (12.7\%)& \textbf{(29.3\%)}\\
\multirow{2}{2.5em}{WT2}&  \multirow{2}{2.5em}{4.93}   &4.66 &  4.32& \textbf{3.69}	\\
& & (5.3\%) &  (12.2\%) & \textbf{ (25.0\%)}\\
\multirow{2}{2.5em}{WT3}  &   \multirow{2}{2.5em}{3.95} & 4.11  &  3.50 & \textbf{3.19}	\\
& & ($-$4.1\%) & (11.4\%) & \textbf{(19.2\%)}\\
\multirow{2}{2.5em}{WT4} &   \multirow{2}{2.5em}{3.73}   & 3.96 &  3.47& \textbf{2.94}	\\
& &  ($-$6.5\%) &  (6.7\%) &  \textbf{(21.0\%)}\\
\hline		
\end{tabular}
\end{table}

\begin{table}[h]
\spacingset{1}
\caption{A comparison table for out-of-temporal RMSE for dataset $ \mc{T}_3 $  } \label{Table:RMSE_T3}
\centering
\begin{tabular}{c | c c c c}
\hline
\textbf{Dataset} & Binning &  kNN & AMK   &  tempGP ($ f(\bs{x}) $) \\
\hline
\multirow{2}{2.5em}{WT1}   & \multirow{2}{2.5em}{5.03}  & 4.98  & 4.47 & \textbf{4.11}\\
& & (0.8\%) & (11.0\%) &  \textbf{(18.1\%)}\\
\multirow{2}{2.5em}{WT2} & \multirow{2}{2.5em}{5.17} & 5.04  &  4.78 & 	\textbf{4.48}\\
& & (2.3\%) & (7.4\%) & \textbf{(13.2\%)}\\
\multirow{2}{2.5em}{WT3}  &    \multirow{2}{2.5em}{4.23}  & 4.68  &  4.12 & \textbf{3.83}\\
& & ($-$10.6\%) & (2.6\%) & \textbf{(9.5\%)}\\
\multirow{2}{2.5em}{WT4} &    \multirow{2}{2.5em}{3.64} & 4.05&3.14  & \textbf{2.84}	\\
& &  ($-$11.3\%) & (13.7\%) &  \textbf{(22.0\%)}\\
\hline		
\end{tabular}
\end{table} 	

Tables \ref{Table:RMSE_T2} and \ref{Table:RMSE_T3} present the performance of binning, kNN, AMK, and tempGP for out-of-temporal predictions on test dataset $ \mc{T}_2 $ and $ \mc{T}_3 $, respectively. Since $ \mc{T}_2 $ and $ \mc{T}_3 $ are temporally disjoint from $ \mc{T}_1 $, we only use the estimate for the time-invariant function $ f(\bs{x}) $ for predictions. We highlight in boldface font the best prediction performance, i.e., whichever has the lowest RMSE. The values in parentheses denote the percentage improvement over the binning method. A negative percentage implies a worse performance than binning. Evidently, tempGP outperforms both the industry baseline (binning) and the data science competitors (kNN and AMK). In other words, explicitly avoiding the temporal structure in the learned function improves the performance of out-of-temporal predictions.
	
Next, we present the results for the second category of methods (TS-kNN, CVc-kNN, and PW-AMK) in Tables \ref{Table:SA_RMSE_T2} and \ref{Table:SA_RMSE_T3}. We also append the results for kNN and AMK from Tables \ref{Table:RMSE_T2} and \ref{Table:RMSE_T3} for easier comparison. The last column of the tables (\% Imp) shows the percentage improvement for tempGP over the second best method; note that the second best method differs for different datasets. For the cross-validation based methods, TS-kNN and CVc-kNN, we notice some improvement in performance as compared to their counterpart kNN, in most of the cases. However, our proposed method still outperforms these methods. The pre-whitening method PW-AMK shows a small but sometimes no improvement over its counterpart AMK. Pre-whitening relies on the autocorrelation in the residuals to modify the response. \cite{Lee2015} show that the autocorrelation in the residuals of the AMK model is still there but nonetheless weakened. As \cite{roberts2017} explained, the temporal structure in the residual can easily get modeled through some input variables when multiple autocorrelated input variables are present, masking the autocorrelation of the residual.  Thus, this weakened autocorrelation in the residual may not be strong enough to modify the response significantly in the pre-whitening step.

Overall, looking at the results of these four datasets, our proposed method is a clear winner. However, the second best method varies from case to case. Interestingly, the standard version of AMK, which does not model the temporal structure, turn out to be the second best in some of the cases. \cite{rabinowicz2020} explain that if the correlation structure in training and test datasets are the same, there is no need for a special treatment of the correlation structure in the data, and the standard methods would perform well. In practice, we do not know the temporal correlation structure in the training and test datasets, and thus cannot guarantee if the correlation pattern would stay the same. Thus, it is a good idea to assume different correlation structure for training and test sets when one is not certain that the correlation structures are the same. This argument becomes more convincing as we extend our case study to a larger set of datasets. There we notice that not handling the temporal structure results in much worse predictions.

\begin{table}
\spacingset{1}
\centering
\caption{A comparison table for out-of-temporal RMSE for dataset $ \mc{T}_2 $ using methods that account for serial autocorrelation.} \label{Table:SA_RMSE_T2}
\begin{tabular}{c | c c c c | c c | c}
\hline
\textbf{Dataset} &  tempGP  & TS-kNN & CVc-kNN & PW-AMK &kNN & AMK &  \% Imp\\
\hline
WT1  & \textbf{3.52} & 4.06 & 4.10 & 4.38 & 4.96 & 4.35 & 13.3\% \\
WT2 & \textbf{3.69}	& 4.59& 4.70&4.38 & 4.66 & 4.32 & 14.6\%\\
WT3  &  \textbf{3.19}	& 3.98& 3.73& 3.39 & 4.11 & 3.50 & 5.8\%\\
WT4 &  \textbf{2.94}	& 3.82& 3.55& 3.47 & 3.96 & 3.47 & 15.2\%\\
\hline			
\end{tabular}
\end{table}

\begin{table}
\spacingset{1}
\centering
\caption{A comparison table for out-of-temporal RMSE for dataset $ \mc{T}_3 $ using methods that account for serial autocorrelation.} \label{Table:SA_RMSE_T3}
\begin{tabular}{c | c c c c | c c | c}
\hline
\textbf{Dataset} &  tempGP  & TS-kNN & CVc-kNN & PW-AMK & kNN & AMK & \% Imp\\
\hline
WT1  & \textbf{4.11} & 4.26& 4.25 &  4.49 & 4.98 & 4.47 & 3.3\%\\
WT2 & \textbf{4.48}	& 5.14& 5.12&4.82 & 5.04 & 4.78 & 6.3\%\\
WT3  &  \textbf{3.83}	& 4.32& 4.11& 3.97 & 4.68& 4.12 & 3.6\%\\
WT4 &  \textbf{2.84}	& 3.74& 3.51& 3.19 & 4.05& 3.14 & 9.6\%\\
\hline		
\end{tabular}
\end{table}

\subsection{Results for Case Study II}\label{Section:Results2}
\begin{figure}
\spacingset{1}
\centering
\includegraphics[width = 0.45\textwidth]{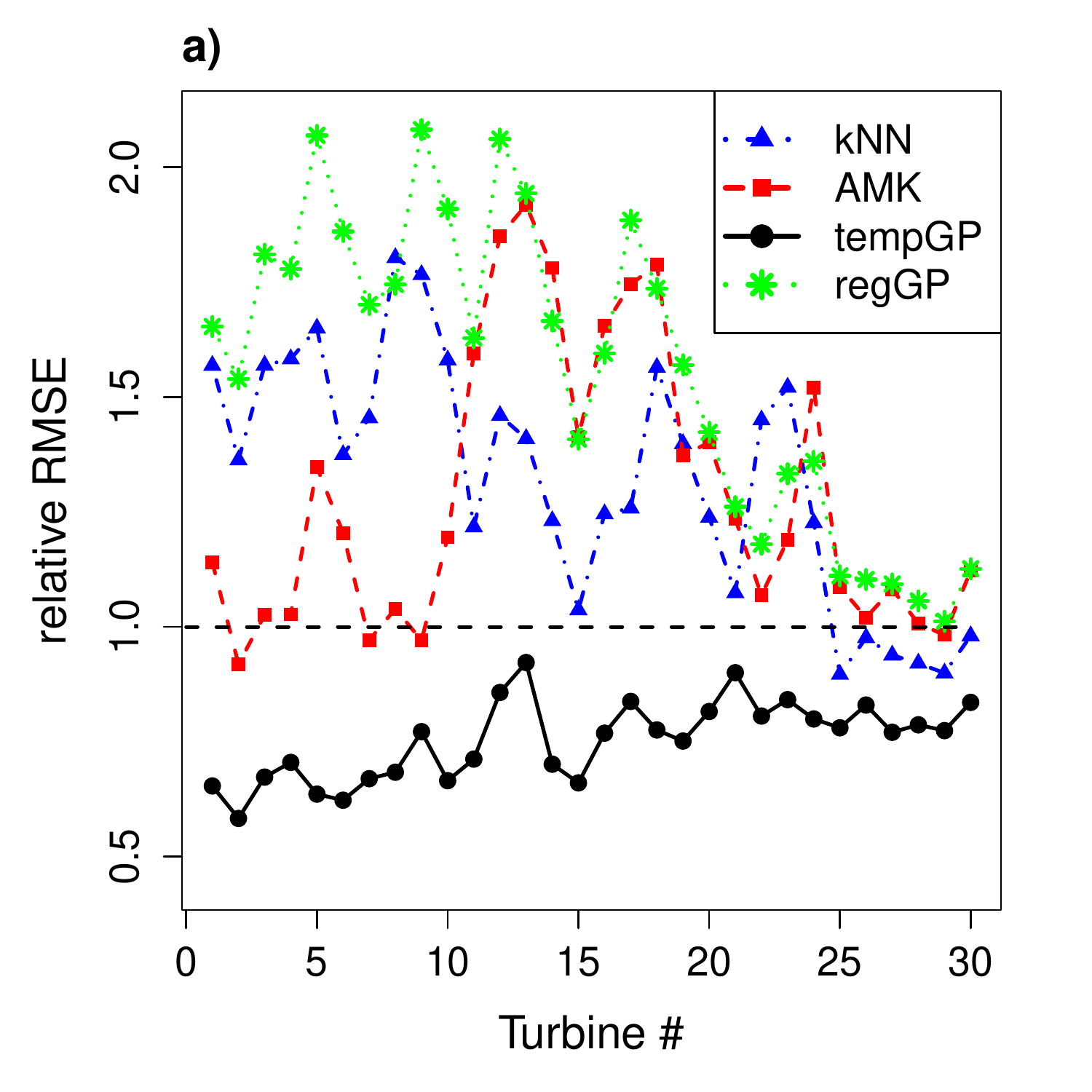} \qquad \includegraphics[width = 0.45\textwidth]{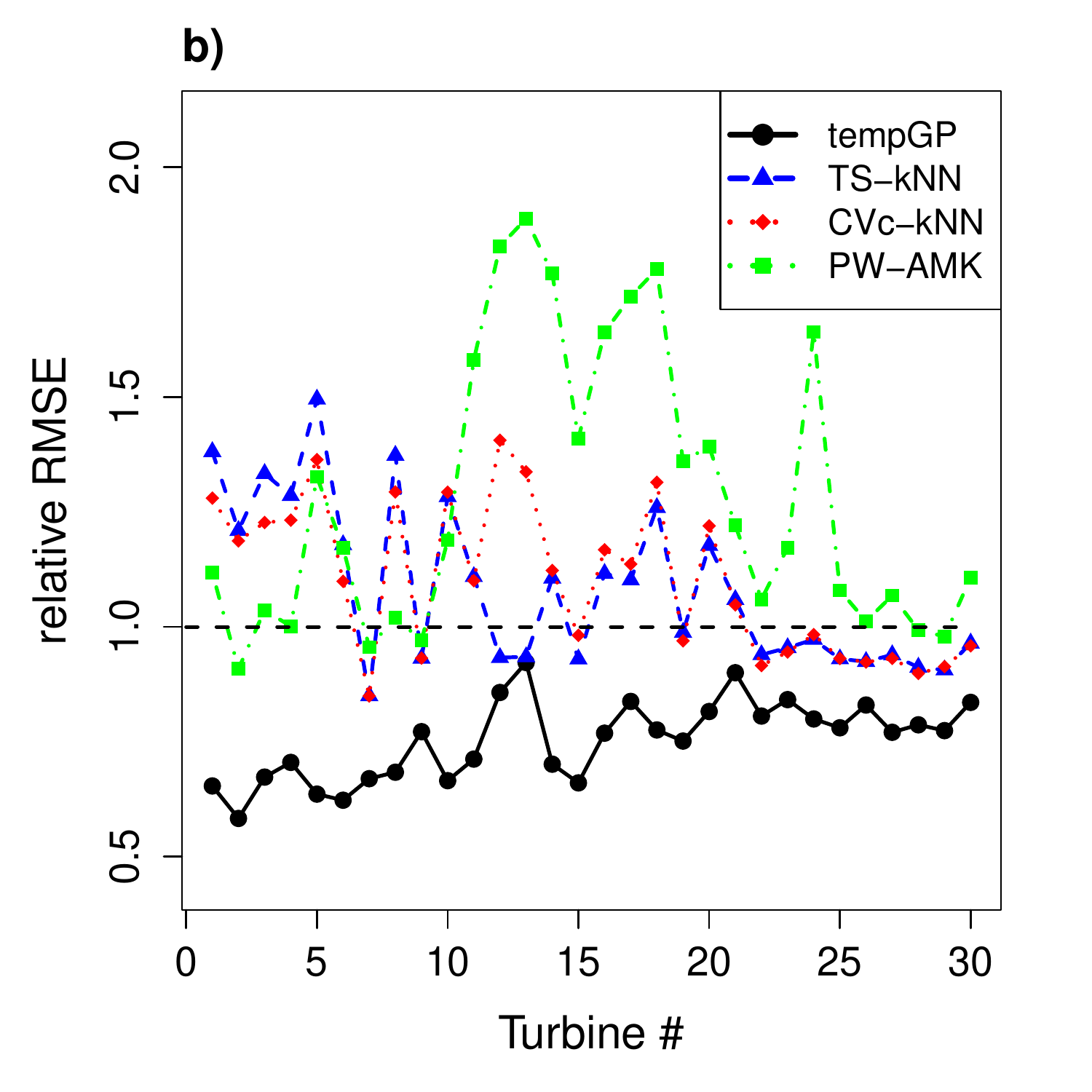} \\
\includegraphics[width = 0.45\textwidth]{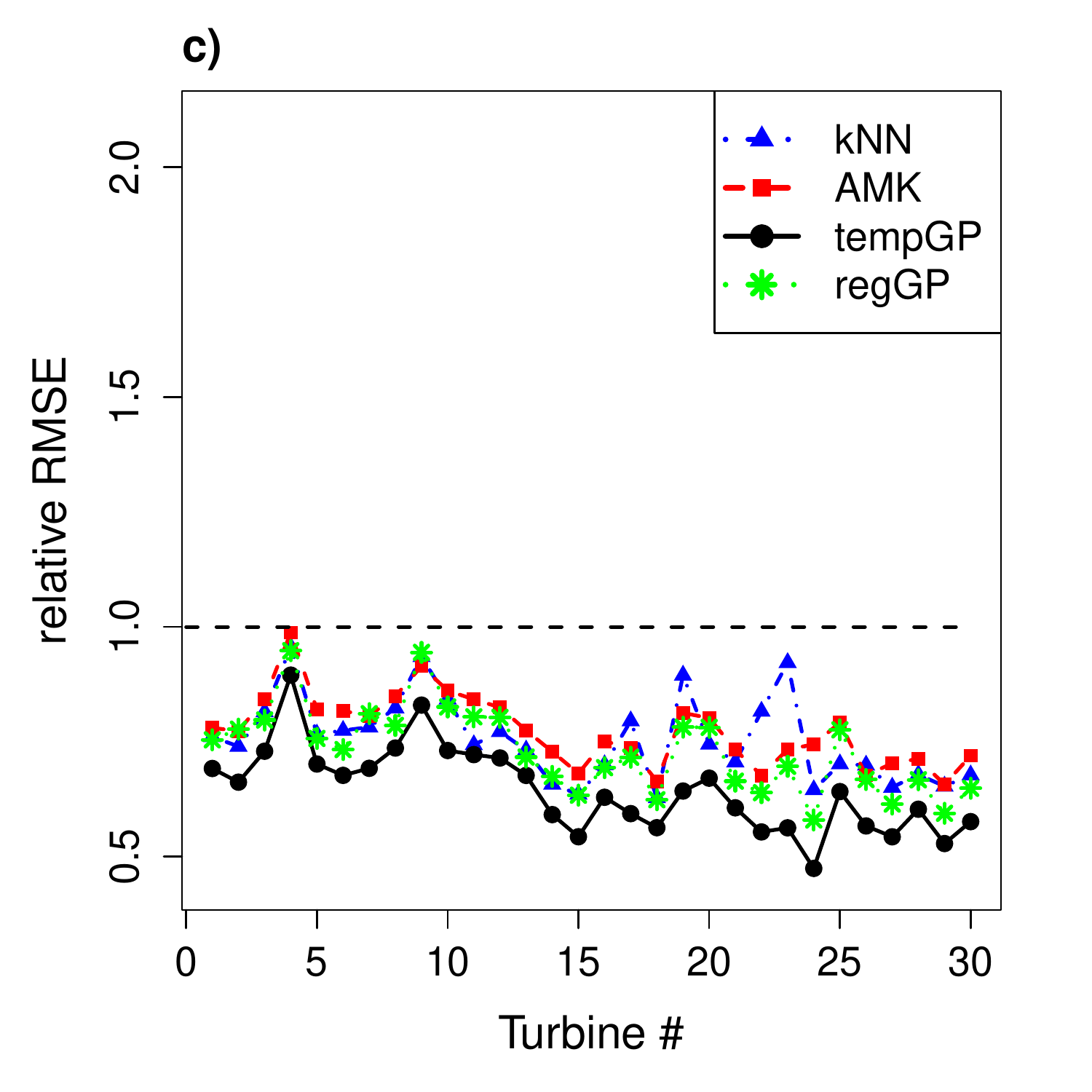} \qquad 	\includegraphics[width = 0.45\textwidth]{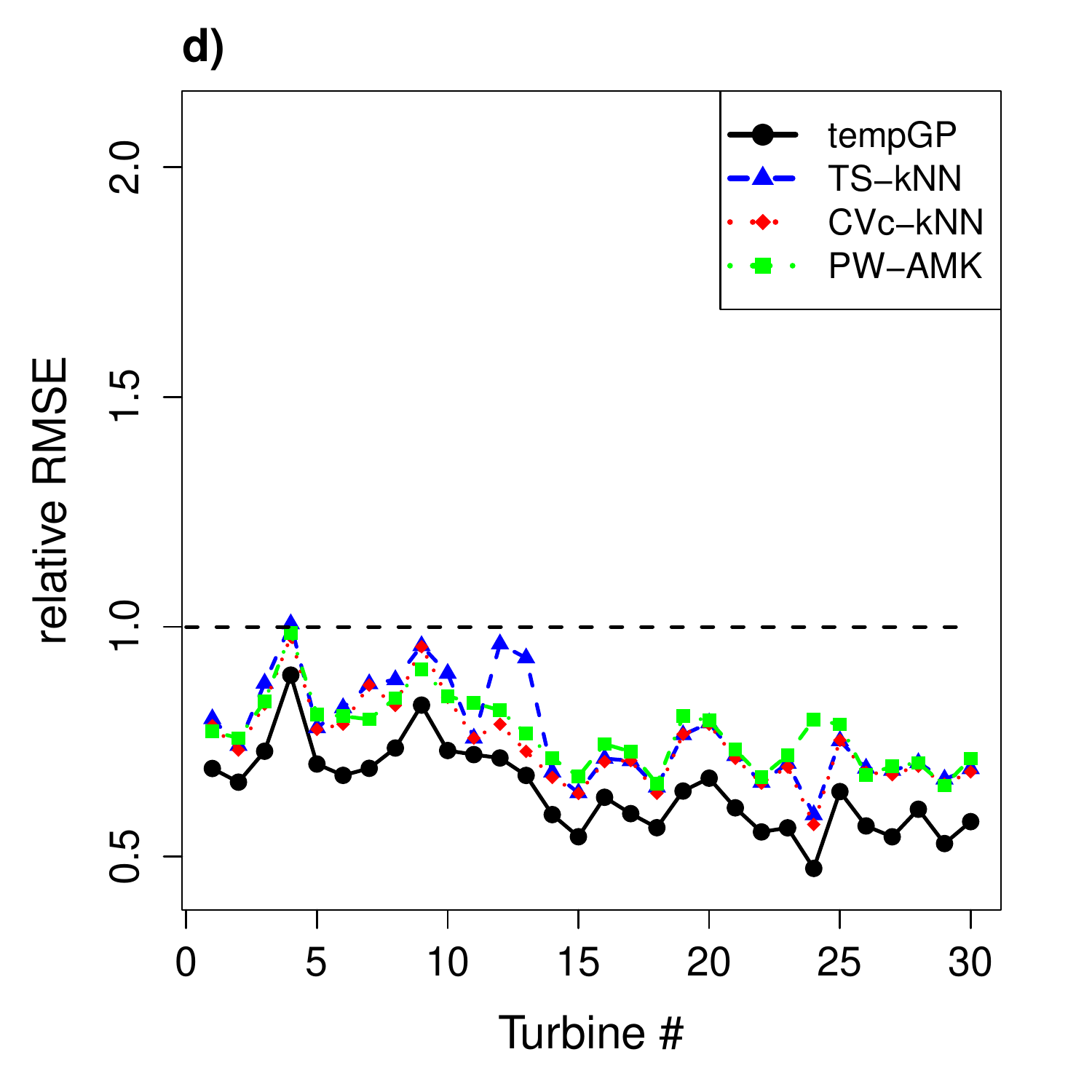}
\caption{Relative RMSEs as compared to binning RMSE for out-of-temporal datasets. The top two plots are for dataset $ \mc{T}_2 $ with the top-left plot a) for kNN, AMK, tempGP, and regGP and the top-right plot b) for TS-kNN, CVc-kNN, PW-AMK, and tempGP. The bottom two plots are for dataset $ \mc{T}_3 $ with the bottom-left plot c) for kNN, AMK, tempGP, and regGP and the bottom-right plot d) or TS-kNN, CVc-kNN, PW-AMK, and tempGP.}\label{Fig:relRMSE}
\end{figure}

We extend our case study on another thirty datasets. Since the datasets in Case Study II are of smaller size, we also used a regular version of Gaussian process, that is, without the $g(t)$ term and thinning, and refer it as regGP. In order to present the results concisely, we use plots instead of the tables. Figure \ref{Fig:relRMSE} (left panels) presents the relative RMSEs of regGP, tempGP, kNN and AMK with respect to binning. To obtain the relative RMSEs, we divide all the RMSEs for different methods by the RMSE of the binning method for each turbine, so that the relative RMSEs in the same scale. Thus, a value larger than one implies performance deterioration over binning; for example, a relative RMSE of 1.1 implies that the method performs 10\% worse than binning. A relative RMSE smaller than one implies performance improvement over binning. The dashed horizontal line at one is for the binning method.  The actual RMSEs are presented in Supplementary Material S4. We also plot the prediction intervals for select turbines in Supplementary Material S5.

We notice that for the first out-of-temporal dataset $ \mc{T}_2 $ (Figure \ref{Fig:relRMSE}a), the performance of kNN, AMK, and, regGP are much worse than that of the second out-of-temporal dataset $ \mc{T}_3$ (Figure \ref{Fig:relRMSE}c). As mentioned earlier, these datasets are for 15 to 18 months time periods. Thus, equal division of these datasets into $ \mc{T}_1 $, $ \mc{T}_2 $, and $ \mc{T}_3 $ results in a half year time span for each of $ \mc{T}_1 $, $ \mc{T}_2 $, and $ \mc{T}_3 $.  What this means is that if $ \mc{T}_1 $ cover the first half of a year, then $ \mc{T}_2 $ covers the second half and $ \mc{T}_3 $ covers the first half of the next year. Due to the seasonal difference of environmental variables, principally that of wind and temperature, it is commonly understood that using the first half year data to predict the second half of the same year is harder than using the first half year data to predict the same first half year of the next year. In another angle, $ \mc{T}_1 $ and $ \mc{T}_2 $ have rather different temporal structure but $ \mc{T}_1 $ and $ \mc{T}_3 $ share a similar temporal structure. It does not therefore come as a surprise that a model temporally-overfitted on $ \mc{T}_1 $ could perform much worse on $ \mc{T}_2$ but reasonably well on $ \mc{T}_3 $. We stress that tempGP performs uniformly better for both out-of-temporal test datasets, although the performance gain is admittedly much more pronounced for $ \mc{T}_2 $.
	
Figure \ref{Fig:relRMSE} (right panels) presents the relative RMSEs (still relative to binning) for the second category of methods---TS-kNN, CVc-kNN, and PW-AMK---along with tempGP. When the temporal overfitting causes a much worse performance for standard regression methods (kNN, AMK, and regGP), which is the case for $ \mc{T}_2 $, the second category of methods that address temporal overfitting provides a significant improvement over their non-temporal counterpart, except for PW-AMK. When temporal overfitting does not result in a worse performance (for $ \mc{T}_3 $), we do not see much help from these temporal methods. It may be fair to say that these temporal methods are not very sensitive to weak temporal overfitting.

\subsection{Further experiments and simulations}\label{Section:ThinExp}

The main parameter used in our method is the thinning number, as it regulates the temporal autocorrelation in each of the data bins. Thus, in order to highlight the importance of thinning and validate our choice of thinning number, we did a sensitivity analysis using different thinning numbers on Case Study II datasets.
We consider the following thinning numbers: $1, 2, 2^2=4, \dots, 2^6=64$. The thinning number computed from our proposed approach for these datasets vary between 14 and 17 with a majority of them being 15. Using these thinning numbers, we re-estimate the function $f$ and recompute the test errors for $\mathcal{T}_2$ and $\mathcal{T}_3$. A thinning number of 1 implies no thinning at all, which is essentially a regular version of GP model without the $g(t)$ term, same as regGP in Case Study II.

\begin{figure}
\centering
\includegraphics[width=0.45\linewidth]{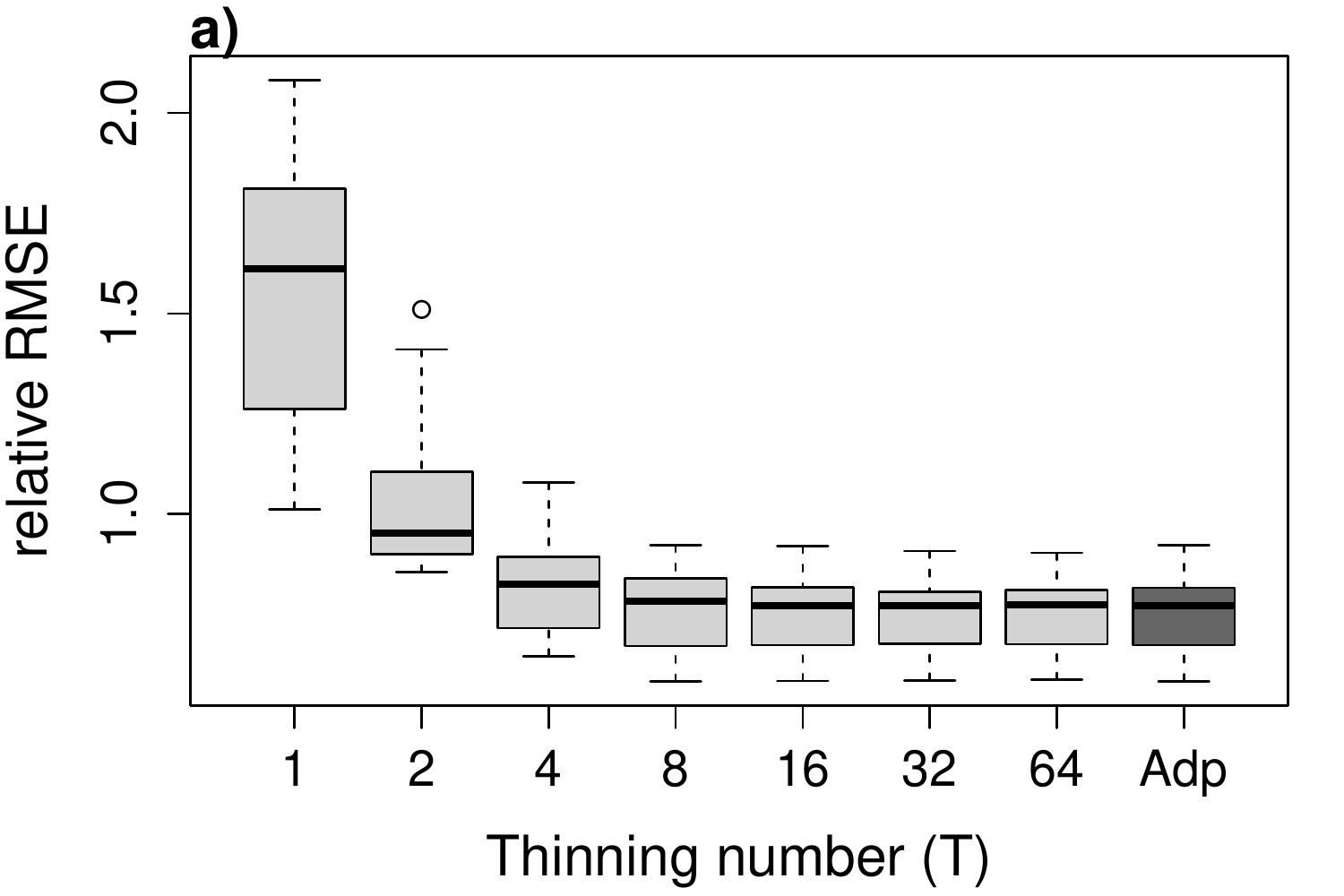} \quad
\includegraphics[width=0.45\linewidth]{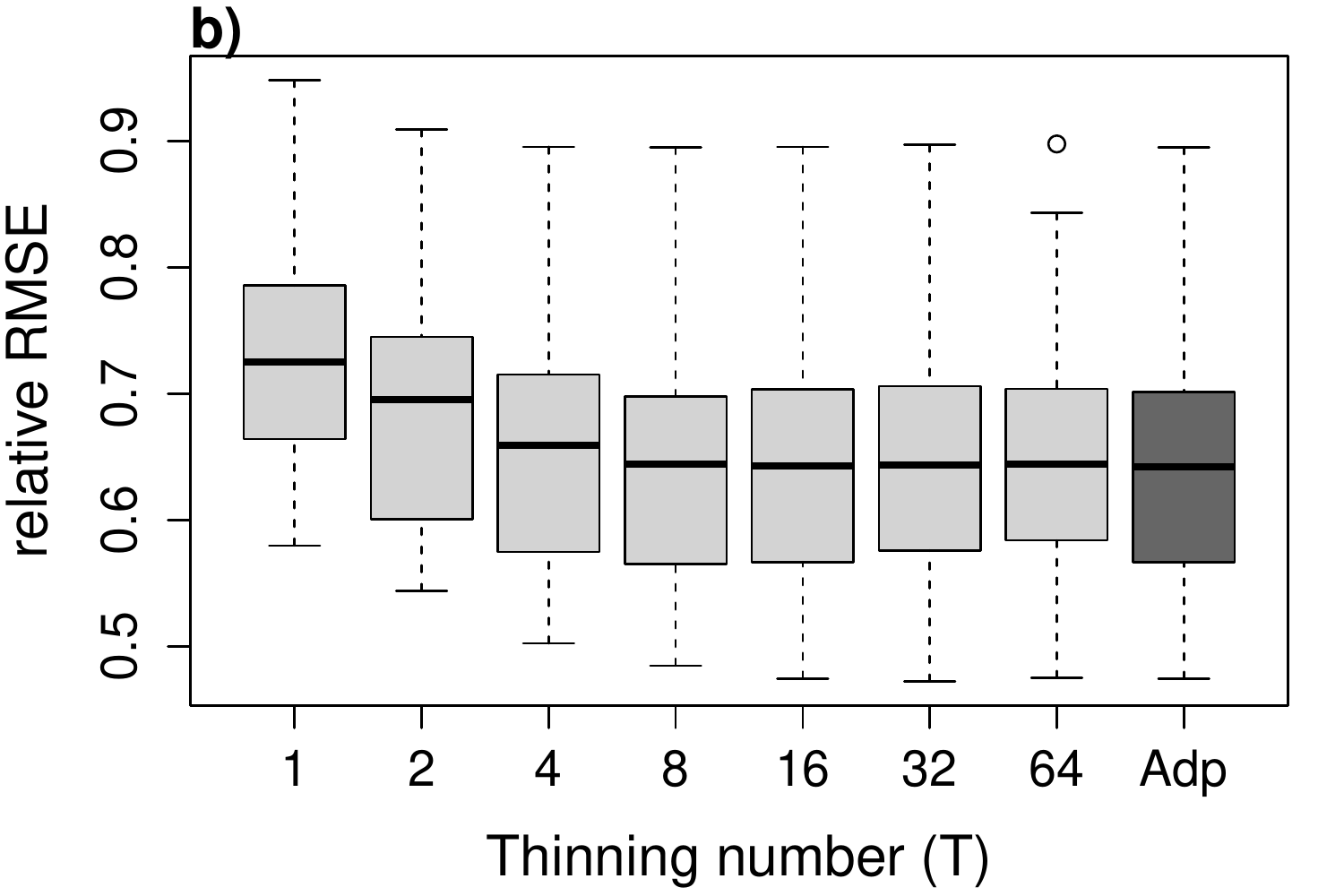}
\caption{Box plots for relative RMSE using different thinning numbers for all the turbines: a) for test set $\mathcal{T}_2$; b) for test set $\mathcal{T}_3$. ``Adp" denotes the adaptive thinning number computed using the proposed approach.} \label{fig:thinNum_boxplots}
\end{figure}

We present the box plots for relative RMSEs (defined in Section \ref{Section:Results2}) for all the turbines in Figure \ref{fig:thinNum_boxplots}. The advantage of thinning is much more pronounced in $\mathcal{T}_2$ than $\mathcal{T}_3$. This is consistent with our previous comments in Section \ref{Section:Results2} about $\mathcal{T}_1$ and $\mathcal{T}_2$ having different temporal structure, and  $\mathcal{T}_1$ and $\mathcal{T}_3$ having similar temporal structure because they are approximately the same time period of two consecutive years. We see that when temporal structure between the training and test datasets are different, thinning plays an important role in improving the performance, and our proposed approach for computing the thinning number, referred to as ``Adp" in the figure, proves to be quite effective.

We also applied our method on a simulated function where the ground truth is known. The details of the simulation study is available in Supplementary Material S6.	

\subsection{Direct inference of $ f(\cdot) $ and $ g(\cdot) $}\label{Section:DirectEst}
\begin{figure}[h]
	\spacingset{1}
	\centering
	\includegraphics[width=0.5\textwidth]{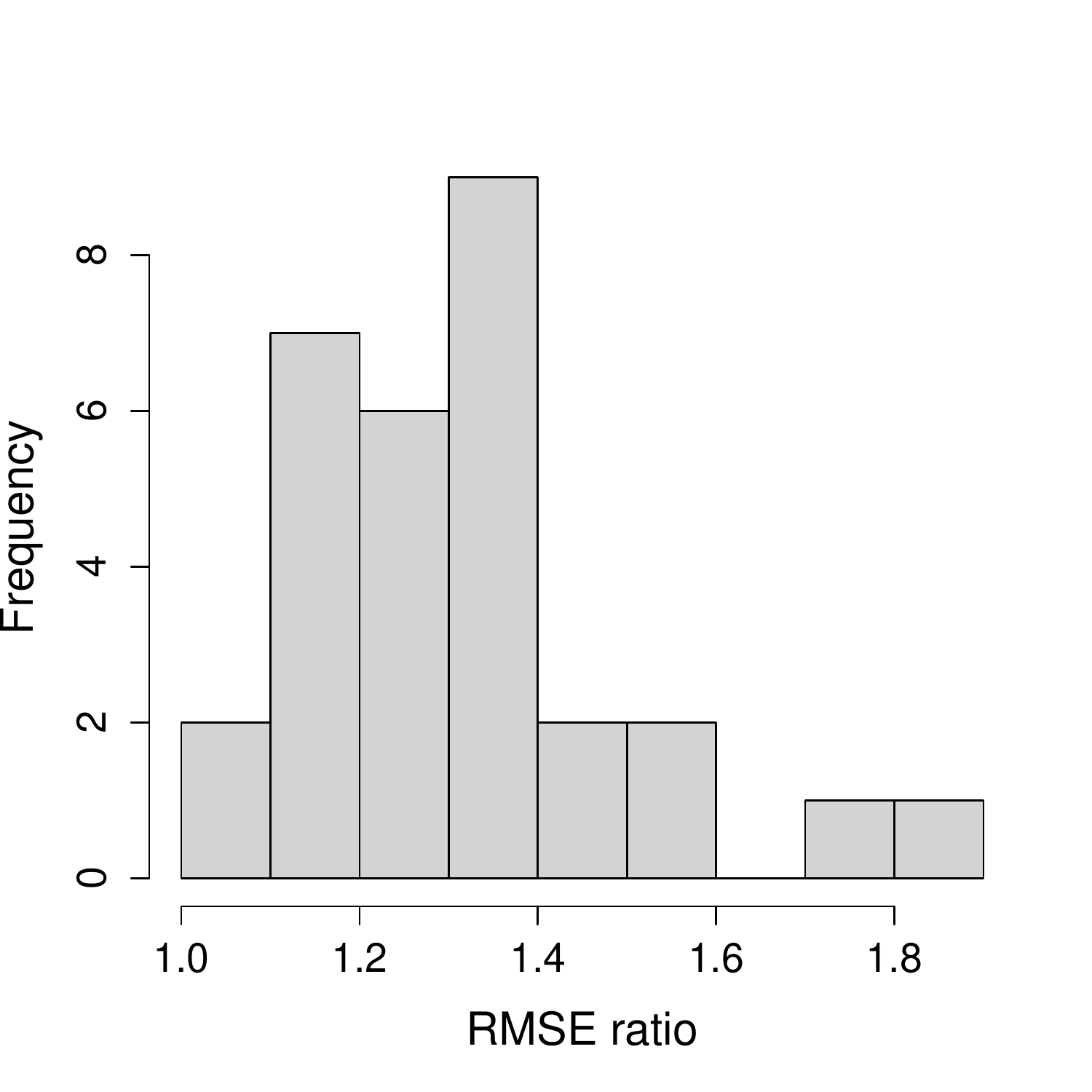}
	\caption{Ratios of the out-of-temporal RMSEs for $\mathcal{T}_2$ obtained from jointly estimated hyperparameters over those using hyperparameters from the thinning-based inference.}\label{Fig:RmseRatio}
\end{figure}

In the model inference section, we state that estimating the hyperparameters of $ f(\cdot) $ and $ g(\cdot) $ jointly via a maximum likelihood estimation results in an identifiability problem, which can result in unreliable hyperparameter estimates and thereby leads to considerable deterioration in prediction performance. We here provide the numerical evidence on the 30-turbine $\mathcal{T}_2$ datasets used in Case Study II. Figure \ref{Fig:RmseRatio} presents the histograms of the ratios of the out-of-temporal RMSEs obtained by using the jointly estimated hyperparameters over that obtained by the thinning-based inference.

We find that under the best case scenario, the direct (joint) estimation results in an error rate that is 6\% worse than that of the thinning-based inference, and under the worst case scenario, the direct estimation results in an error rate that is 80\% worse, that is the ratio of out-of-temporal RMSE for direct estimation vs thinning-based estimation is approximately 1.8. Out of the 30 turbines, 28 cases are at least 10\% worse.

Another advantage of thinning-based approach over the direct estimation is the computation time.  In general, the time complexity for fitting a GP model is of the order $\mathcal{O}(n^3)$, where $n$ is the number of data points. However, since we bin the data into $T$ bins such that each bin has approximately $n/T = m$ observations and use the pseudo-likelihood defined in Section \ref{Section:Inference}, the time complexity for fitting the proposed model is in the order of $\mathcal{O}(Tm^3) = \mathcal{O}(nm^2)$, which is lower than $\mathcal{O}(n^3)$ for any $T > 1$.

\subsection{Further discussions on CVc}\label{Section:CVc}
Looking at all methods that account for correlation in data, CVc looks like the most promising alternative to tempGP. CVc, however, has its own limitations.

The main challenge that arises in our problem setting is the estimation of $ Cov(\bs{y},\bs{y}) $, which plays a critical role in correcting the standard CV error estimation. As noted in Section \ref{Section:Method}, \cite{rabinowicz2020} assume the input variables to be i.i.d. Under their assumption, $ Cov(\bs{y},\bs{y}) $ can be directly estimated using sample covariance matrix of $\bs{y}$ because no covariance in $ \bs{y} $ is due to the input variables in $\bs{x}$. For our problem setting, however,  $ \bs{x} $ are autocorrelated. We need to estimate the covariance in $ \bs{y} $ due to $g(t)$. This dual autocorrelation makes it harder to estimate $ Cov(\bs{y},\bs{y}) $ accurately. Strictly speaking, CVc as presented in \cite{rabinowicz2020} is not directly applicable to our model.

Our \emph{ad hoc} procedure in Section~\ref{Section:Implementation} is an attempt to estimate $ Cov(\bs{y},\bs{y}) $ in the presence of the dual autocorrelation. We acknowledge that the \emph{ad hoc} procedure may not be the best approach, but it remains unknown how to estimate the covariance in $ \bs{y} $ due to $g(t)$ under our problem setting. While devising the \emph{ad hoc} procedure, we used the residuals of a fixed one-dimensional kNN model to estimate  $ Cov(\bs{y},\bs{y}) $. One may ask if it would be better to increase the number of input variables in the kNN model while estimating $ Cov(\bs{y},\bs{y})$? Using a multivariate model weakens the correlation in the residuals, leading to a different estimate of  $ Cov(\bs{y},\bs{y}) $. As such, the issue of variable selection gets entangled with the estimate of CVc. It is unclear to us which multivariate model should be used for estimating $ Cov(\bs{y},\bs{y})$. Given these challenges with CVc, tempGP appears better suited for the application at hand. The empirical evidence is rather strong in supporting this claim.

\section{Conclusion}\label{Section:Disc}
We explore a class of regression problems when the input variables and errors are serially correlated over time. Classical regression, which works under the independence assumption, results in overfitted models, known as temporal overfitting.  We propose a method to reduce temporal overfitting by explicitly modeling the temporal correlation in the data. We split the variance in response into a time-independent function and a temporally autocorrelated stochastic process. We take advantage of an idea frequently used in Bayesian statistics---thinning. Using the thinned data, the time-independent function can be separately estimated from the temporally autocorrelated model term.

The thinning-based idea is one of the approaches that can be used to learn the time-independent function. An alternative approach could be to regularize the time-independent function $f$, or constrain it, following an idea first proposed in \cite{Ba2012}.  \citet{Ba2012} also considers an additive model with two GP terms. They separate the effect of the two terms by ensuring that one term is smoother than the other and then constraining the lengthscale of the two kernels accordingly.  Unlike in \cite{Ba2012} where the two GP terms take the same input, $f$ and $g$ in our model take different inputs, and as a result, it is not immediately clear how the lengthscales of the respective kernels should be constrained, but this could be an interesting future work to pursue.
	
A final note is that while the paper highlights the problem of temporal overfitting in wind power curves, we believe that the wind energy problem is  just one of the many application areas where one could encounter temporal overfitting. Many real datasets in engineering and life sciences are collected over time and could be autocorrelated due to the inertia in the underlying physical processes. We are confident that the resulting methodology is generic and could benefit other nonparametric regressions of the same nature.

\if0\blind{
\vspace{-12 pt}
\section*{Supplementary Material}
\begin{description}
\item[Supplementary Material:] The PDF file contains: (S1) Results for Case Study II with different covariance functions, (S2) PACF plots for WT1, (S3) Hyperparameter estimates, (S4) Actual RMSEs for Case Study II, (S5) Prediction intervals for select turbines, and (S6) Experiments on a simulated function.
\item[Computer Code:] The computer code to reproduce all the results in this paper are available on \texttt{GitHub} at \url{https://github.com/TAMU-AML/tempGP-Paper}. A generic \texttt{R} function for applying the tempGP algorithm to any dataset is available in \texttt{DSWE} package in \texttt{R} available through CRAN at \url{https://CRAN.R-project.org/package=DSWE}.
\vspace{-12 pt}
\end{description}

\vspace{-12 pt}
\section*{Funding}
Prakash and Ding's research is partially supported by NSF IIS-1741173; Tuo's research by NSF DMS-1914636; and Ding and Tuo's research also by NSF grant CCF-1934904.
}\fi

\vspace{-3 pt}

\bibliographystyle{apalike}
\spacingset{1}
\bibliography{TemporalOverfitting}

\begin{thebibliography}{}

\bibitem[Altman, 1990]{Altman1990}
Altman, N.~S. (1990).
\newblock {Kernel smoothing of data with correlated errors}.
\newblock {\em Journal of the American Statistical Association},
  85(411):749--759.

\bibitem[Ba and Joseph, 2012]{Ba2012}
Ba, S. and Joseph, V.~R. (2012).
\newblock Composite {G}aussian process models for emulating expensive
  functions.
\newblock {\em The Annals of Applied Statistics}, 6(4):1838--1860.

\bibitem[Bessa et~al., 2012]{bessa2012time}
Bessa, R.~J., Miranda, V., Botterud, A., Wang, J., and Constantinescu, E.~M.
  (2012).
\newblock Time adaptive conditional kernel density estimation for wind power
  forecasting.
\newblock {\em IEEE Transactions on Sustainable Energy}, 3(4):660--669.

\bibitem[Burman et~al., 1994]{burman1994}
Burman, P., Chow, E., and Nolan, D. (1994).
\newblock A cross-validatory method for dependent data.
\newblock {\em Biometrika}, 81(2):351--358.

\bibitem[Chipman et~al., 2010]{Chipman2010}
Chipman, H.~A., George, E.~I., and McCulloch, R.~E. (2010).
\newblock {BART}: Bayesian additive regression trees.
\newblock {\em The Annals of Applied Statistics}, 4(1):266--298.

\bibitem[Chu and Marron, 1991]{chu1991}
Chu, C.-K. and Marron, J.~S. (1991).
\newblock Comparison of two bandwidth selectors with dependent errors.
\newblock {\em The Annals of Statistics}, 19(4):1906--1918.

\bibitem[De~Brabanter et~al., 2018]{debrabanter2018}
De~Brabanter, K., Cao, F., Gijbels, I., and Opsomer, J. (2018).
\newblock Local polynomial regression with correlated errors in random design
  and unknown correlation structure.
\newblock {\em Biometrika}, 105(3):681--690.

\bibitem[De~Brabanter et~al., 2011]{debrabanter2011}
De~Brabanter, K., De~Brabanter, J., Suykens, J.~A., and De~Moor, B. (2011).
\newblock Kernel regression in the presence of correlated errors.
\newblock {\em Journal of Machine Learning Research}, 12(55):1955--1976.

\bibitem[Ding, 2019]{Ding2019}
Ding, Y. (2019).
\newblock {\em Data Science for Wind Energy}.
\newblock Chapman \& Hall, Boca Raton, FL.

\bibitem[EIA, 2021]{eia}
EIA (2021).
\newblock {US Energy Information Administration}.
\newblock Webpage:
  \url{https://www.eia.gov/energyexplained/wind/electricity-generation-from-wind.php},
  Accessed on: 2021-09-17.

\bibitem[Geller and Neumann, 2018]{geller2018}
Geller, J. and Neumann, M.~H. (2018).
\newblock Improved local polynomial estimation in time series regression.
\newblock {\em Journal of Nonparametric Statistics}, 30(1):1--27.

\bibitem[Gu, 2013]{Gu2013Book}
Gu, C. (2013).
\newblock {\em Smoothing Spline ANOVA Models}.
\newblock Springer-Verlag, New York, 2nd edition.

\bibitem[Gu, 2014]{Gu2014Package}
Gu, C. (2014).
\newblock Smoothing spline {ANOVA} models: R package gss.
\newblock {\em Journal of Statistical Software, Articles}, 58(5):1--25.

\bibitem[Hastie et~al., 2009]{hastie2009}
Hastie, T., Tibshirani, R., and Friedman, J. (2009).
\newblock {\em The Elements of Statistical Learning: Data Mining, Inference,
  and Prediction}.
\newblock Springer, New York, NY, 2nd edition.

\bibitem[IEC, 2005]{IEC05}
IEC (2005).
\newblock {\em IEC TS 61400-12-1 Ed. 1, Power Performance Measurements of
  Electricity Producing Wind Turbines}.
\newblock Geneva, Switzerland.

\bibitem[Kumar et~al., 2021]{DSWE-Package}
Kumar, N., Prakash, A., and Ding, Y. (2021).
\newblock {\em DSWE: Data Science for Wind Energy}.
\newblock R package version 1.5.1,
  \url{https://CRAN.R-project.org/package=DSWE}.

\bibitem[Lee et~al., 2015]{Lee2015}
Lee, G., Ding, Y., Genton, M.~G., and Xie, L. (2015).
\newblock Power curve estimation with multivariate environmental factors for
  inland and offshore wind farms.
\newblock {\em Journal of the American Statistical Association},
  110(509):56--67.

\bibitem[Meyer et~al., 2018]{meyer2018}
Meyer, H., Reudenbach, C., Hengl, T., Katurji, M., and Nauss, T. (2018).
\newblock Improving performance of spatio-temporal machine learning models
  using forward feature selection and target-oriented validation.
\newblock {\em Environmental Modelling \& Software}, 101:1--9.

\bibitem[Opsomer et~al., 2001]{Opsomer2001}
Opsomer, J., Wang, Y., and Yang, Y. (2001).
\newblock Nonparametric regression with correlated errors.
\newblock {\em Statistical Science}, 16(2):134--153.

\bibitem[Rabinowicz and Rosset, 2020]{rabinowicz2020}
Rabinowicz, A. and Rosset, S. (2020).
\newblock Cross-validation for correlated data.
\newblock {\em Journal of the American Statistical Association}, in press and
  online available, DOI: 10.1080/01621459.2020.1801451.

\bibitem[Racine, 2000]{racine2000}
Racine, J. (2000).
\newblock Consistent cross-validatory model-selection for dependent data:
  hv-block cross-validation.
\newblock {\em Journal of Econometrics}, 99(1):39--61.

\bibitem[Rasmussen and Williams, 2006]{Rasmussen2006}
Rasmussen, C.~E. and Williams, C. K.~I. (2006).
\newblock {\em Gaussian Processes for Machine Learning}.
\newblock The MIT Press, Cambridge, MA.

\bibitem[Roberts et~al., 2017]{roberts2017}
Roberts, D.~R., Bahn, V., Ciuti, S., Boyce, M.~S., Elith, J., Guillera-Arroita,
  G., Hauenstein, S., Lahoz-Monfort, J.~J., Schröder, B., Thuiller, W.,
  Warton, D.~I., Wintle, B.~A., Hartig, F., and Dormann, C.~F. (2017).
\newblock Cross-validation strategies for data with temporal, spatial,
  hierarchical, or phylogenetic structure.
\newblock {\em Ecography}, 40(8):913--929.

\bibitem[Ruppert et~al., 1995]{ruppert1995}
Ruppert, D., Sheather, S.~J., and Wand, M.~P. (1995).
\newblock An effective bandwidth selector for local least squares regression.
\newblock {\em Journal of the American Statistical Association},
  90(432):1257--1270.

\bibitem[Sheridan, 2013]{sheridan2013}
Sheridan, R.~P. (2013).
\newblock Time-split cross-validation as a method for estimating the goodness
  of prospective prediction.
\newblock {\em Journal of chemical information and modeling}, 53(4):783--790.

\bibitem[Tuo and Wu, 2016]{tuo2016theoretical}
Tuo, R. and Wu, C.~J. (2016).
\newblock A theoretical framework for calibration in computer models:
  {P}arametrization, estimation and convergence properties.
\newblock {\em SIAM/ASA Journal on Uncertainty Quantification}, 4(1):767--795.

\bibitem[Tuo and Wu, 2018]{tuo2018prediction}
Tuo, R. and Wu, C.~J. (2018).
\newblock Prediction based on the {K}ennedy-{O}'{H}agan calibration model:
  {A}symptotic consistency and other properties.
\newblock {\em Statistica Sinica}, 8(2):743--759.

\bibitem[Vapnik, 2000]{Vapnik2000}
Vapnik, V. (2000).
\newblock {\em The Nature of Statistical Learning Theory}.
\newblock Springer-Verlag, New York, 2nd edition.

\bibitem[Wang et~al., 2020]{wang2020prediction}
Wang, W., Tuo, R., and Wu, C.~J. (2020).
\newblock On prediction properties of kriging: {U}niform error bounds and
  robustness.
\newblock {\em Journal of the American Statistical Association},
  115(530):920--930.

\bibitem[Xiao et~al., 2003]{xiao2003}
Xiao, Z., Linton, O.~B., Carroll, R.~J., and Mammen, E. (2003).
\newblock More efficient local polynomial estimation in nonparametric
  regression with autocorrelated errors.
\newblock {\em Journal of the American Statistical Association},
  98(464):980--992.

\end{thebibliography}

\end{document}